\documentclass[useAMS,usenatbib]{mnras}

\usepackage{graphicx}
\usepackage{standalone}
\usepackage{subfig}
\usepackage{multirow}
\usepackage{array}

\newcommand{\xmm}{{\em XMM-Newton}}

\def\lapp{\ifmmode\stackrel{<}{_{\sim}}\else$\stackrel{<}{_{\sim}}$\fi}

\def\nat{Nature}%
\def\apjl{ApJ}%
\def\aap{A\&A}%
\def\physrep{Phys.~Rep.}%

\def\Msun{\rm M_{\odot}}
\def\Av{\rm A_V}
\def\mcom{\rm M_{\rm C}}

\def\0614{\rm PSR\ J0614$-$3329}
\def\1231{\rm PSR\ J1231$-$1411}
\def\2017{\rm PSR\ J2017+0603}

\title[Gemini optical observations of binary millisecond-pulsars]{Gemini optical observations of binary millisecond-pulsars }
\author[V. Testa, R. P. Mignani,   C. Pallanca, A. Corongiu, F. R. Ferraro]
{\parbox{\textwidth}{V. Testa$^{1}$\thanks{E-mail vincenzo.testa@oa-roma.inaf.it}, 
R. P. Mignani$^{2,3}$,
C. Pallanca$^{4}$,
A. Corongiu$^{5}$,
F. R. Ferraro$^{4}$
}
\\ \\
$^{1}$ INAF - Osservatorio Astronomico di Roma, via Frascati 33, 00040, Monteporzio, Italy \\
$^{2}$ INAF - Istituto di Astrofisica Spaziale e Fisica Cosmica Milano, via E. Bassini 15, 20133, Milano, Italy\\
$^{3}$ Kepler Institute of Astronomy, University of Zielona G\'ora, Lubuska 2, 65-265, Zielona G\'ora, Poland \\
$^{4}$ Dipartimento di Fisica e Astronomia, Universit\`a degli Studi di Bologna, Viale Berti Pichat 6-2, I-40127, Bologna, Italy \\
$^{5}$ INAF-Osservatorio Astronomico di Cagliari, Via della Scienza 5, 09047 Selargius, Italy 
}

\begin{document}

\date{Accepted 2015 August 17; Received 2015 August 16; in original form 2015 April 30}

\pagerange{\pageref{firstpage}--\pageref{lastpage}} \pubyear{2015}

\maketitle

\label{firstpage}

\begin{abstract}
Milli-second pulsars (MSPs) are rapidly spinning neutron stars, with spin periods $P_{\rm s} \la 10$ ms,
which have been most likely spun up after a phase of matter accretion from a companion star. 
In this work we present the results of the search for the companion stars of four binary milli-second pulsars, 
carried out with archival data from the Gemini South telescope. Based upon a very good positional coincidence with the pulsar radio coordinates, we likely identified the companion stars to three MSPs, namely PSR\, J0614$-$3329 (g=21.95$\pm$0.05), J1231$-$1411 (g=25.40$\pm$0.23), and J2017+0603 (g=24.72$\pm$0.28). For the last pulsar (PSR\, J0613$-$0200) the identification was hampered by the presence of a bright star (g=16$\pm$0.03) at $\sim$2\arcsec\ from the pulsar radio coordinates and we could only set $3 \sigma$ upper limits of $g=25.0$, $r= 24.3$, and $i= 24.2$ on the magnitudes of its companion star.
The candidate companion stars to PSR\, J0614$-$3329, J1231$-$1411, and  J2017+0603  can be tentatively identified as He white dwarfs (WDs) on the basis of their optical colours and brightness and the comparison with stellar model tracks. From the comparison of our multi-band photometry with stellar model tracks we also obtained possible ranges on the mass, temperature, and gravity of the candidate WD companions to these three MSPs. Optical spectroscopy observations are needed to confirm their possible classification  as He WDs and accurately measure their stellar parameters.

\end{abstract}

\begin{keywords}
stars  neutron -- pulsars  general
\end{keywords}

\section{Introduction}

Radio pulsars are  interpreted as rapidly spinning  and   strongly magnetised neutron stars powered by their rotational   energy.  The most recent compilation  (1.52) of the Australia National Telescope Facility (ATNF) Pulsar Catalogue (Manchester et al.\ 2005) lists  over 2300 radio pulsars. 
%
An important  subgroup ($\sim 270$ objects) is that of the so-called milli-second pulsars (MSPs), characterised by very short spin periods ($P_{\rm s} \la 10$ ms), high  spin stability ($\dot {P_{\rm s}} \approx   10^{-18}$--$10^{-21}$  s~s$^{-1}$), spin-down   ages $P_{\rm s}/2\dot{P_{\rm s}} \sim 1$--10 Gyrs, surface magnetic fields  $B \sim 10^{8}$--$10^{9}$ G, and spin-down energy  $\dot{E} \sim 10^{32}$--$10^{36}$ erg s$^{-1}$.

A large fraction of MSPs (164) are in binary systems and are mostly  located in the Galactic plane.
The fact that many MSPs are found in binary systems supports the commonly accepted scenario for the formation and evolution of MSPs \citep[the ``{\it canonical recycling scenario}'';][]{alpar82, bhattvan91}. This suggests that MSPs form in binary systems from old (slowly rotating) pulsars and that, at a later stage, they are reaccelerated to milli-second spin periods as the result of mass accretion from their non-degenerate companion stars. The pathways to the MSPs are different according to the initial mass ratio and orbital separation of the components, which drives the length of the accretion processes (see, e.g.  Tauris  2011, 2015; Tauris et al.\ 2012).

The most common pathway leads to a companion star that has almost or completely lost its external layers (white dwarf; WD) orbiting a rapidly spinning pulsar \citep[]{lyne87,alpar82,bhattvan91,handbook}.  In  this case, the companion star is expected to be either an He WD of mass  $0.1 M_{\odot} \la M_{\rm C} \la 0.5 M_{\odot}$ or a more massive Carbon-Oxygen (CO) WD, of mass  $0.5 M_{\odot} \la M_{\rm C} \la 1 M_{\odot}$.  When the companion star has not reached the WD stage yet it may be ablated by irradiation from the pulsar relativistic wind, which possibly leads to the formation of a solitary MSP.   Two distinct families of binary MSPs with non-degenerate companions emerge according to the degree of the ablation processes
(e.g., Roberts 2013).  The first one is that of the so-called black widow (BW) MSPs, where the companion is a very low-mass star of $M_{\rm C} \la 0.1 M_{\odot}$, almost fully ablated by the pulsar wind. The second one is that of the redbacks (RB) MSPs, where the companion is only partially ablated and has an higher mass of $M_{\rm C}\sim0.1$--$0.4 M_{\odot}$.
Therefore,  the study of MSPs in binary systems is fundamental to understand the final stages of the binary pulsar evolution, including the different outcomes of the recycling process (e.g., Possenti \& Burgay 2008; Possenti 2013), 
and investigate the possible evolutionary connections between different MSP types  (e.g., Benvenuto et al.\ 2014).

In  all these types of studies, the optical identification of the MSP companion star plays a crucial role (e.g., Pallanca et al.\ 2012, 2013, 2014; Mucciarelli et al.\ 2013). In particular, optical observations, 
 either via spectroscopy or broad-band photometry, are key to determine the star classification and physical parameters, such as surface temperature, surface gravity, and chemical composition of the atmosphere. This information then allow one to determine the age of the binary system (and that of the MSP independently of the measurements on $P_{\rm s}$ and $\dot {P_{\rm s}}$) and track the system evolutionary history. Optical observations also allow one to find evidence for irradiation of the companion star through the identification of hot spots on the stellar surface. This can be a further tracer of the ablation process  in BW and RB systems, 
 after the more direct one which comes from the observation of eclipses 
 of the radio signal as it propagates through the stellar wind from the irradiated MSP companion.
For only $\sim$ 50 binary MSPs the companion star has been identified in the optical. 
Thus, for only about one third of 
the MSPs we know the optical characteristics of the companion star.  In most cases, this is mainly due to the relatively large distances of these systems and the interstellar extinction in the Galactic plane, which hampers deep multi-band optical observations.
 For the vast majority of non-eclipsing MSPs,
 the companion stars are found to be He WDs, although more cases of MSPs with CO WD companions have been found in the last few years (e.g., Mignani et al.\ 2014).  
 

In this manuscript we report on the search for the companion stars of four unidentified binary MSPs carried out using archival data from the Gemini-South Telescope. We describe the MSPs sample and the observations in Section 2, while we present and discuss the results in Section 3 and 4, respectively.

\begin{table*}
\begin{center}
\caption{Coordinates, spin period $P_{\rm s}$ and period derivative $\dot{P_{\rm s}}$ of the observed MSPs, together with the inferred values of the spin-down age ($\tau$), dipolar magnetic field ($B$) and rotational energy loss ($\dot{E}$), as derived from the ATNF pulsar data base.  For the MSP coordinates, the number in parentheses represent the $1 \sigma$ uncertainty on the last quoted digit. For those pulsars for which a proper motion has been measured (PSR\, J0613$-$0200 and J1231$-$1411),
the values of $\dot{P_{\rm s}}$, $\tau$, $B$ and $\dot{E}$, have been corrected for the Shlowskhi effect.}
\label{psr}
\begin{tabular}{lllccccc} \hline
Pulsar 				&$\alpha_{J2000} $	& $\delta_{J2000}$&    P$_{\rm s}$	&    $\dot{P_{\rm s}}$ 	& $\tau$ & B & $\dot{E}$ \\ 
                                          &    $^{(hms)}$   &      $^{(\circ ~'~")}$   &   (ms) &  (10$^{-18}$s s$^{-1}$)    &  ($10^{9}$ yr) & ($10^{8}$ G)  & ($10^{34}$ erg cm$^{-2} s^{-1}$)  \\ \hline
J0613$-$0200$^1$   	&  06 13 43.97514(1.1)      	& $-$02 00 47.1737(4)    	 	  &  3.06 	& 0.00882 	 	 &   5.50 	 &1.66	&  1.3  \\
J0614$-$3329           & 06 14 10.3478(3)                   &$-$33 29 54.118(3)                    &  3.14   & 0.0175                    & 2.84      & 2.38      &  2.2\\
J1231$-$1411$^2$            & 12 31 11.3132(7)                  &$-$14 11 43.63(2)                       & 3.68 & 0.0228                       & 2.56     & 2.93      & 1.8  \\
J2017+0603              & 20 17 22.7044(1)                   &+06 03 05.569(4)                       & 2.89  & 0.0083                       & 5.53    & 1.57      & 1.3 \\ \hline
\end{tabular}
\end{center}
$^1$ $\mu_{\alpha} cos(\delta) =1.84 \pm 0.08$ mas yr$^{-1}$; $\mu_{\delta}=-10.6\pm0.2$ mas yr$^{-1}$ (Verbiest et al.\ 2009) \\
$^2$ $\mu_{\alpha} cos(\delta) =-100 \pm 2$ mas yr$^{-1}$;      $\mu_{\delta}=-30\pm4$ mas yr$^{-1}$ (Ransom et al.\ 2011) 
\end{table*}

\section{Observations}

\subsection{Target Description}

The names, measured radio coordinates, spin period ($P_{\rm s}$), period derivative ($\dot{P_{\rm s}}$) and inferred characteristics, spin-down age ($\tau$), dipolar magnetic field ($B$), spin-down energy ($\dot{E}$), of the four MSPs studied in this work are summarised in Table \ref{psr}. 

Besides the radio band, all these MSPs have been detected as $\gamma$-ray pulsars by {\rm Fermi}  (Abdo et al.\ 2013). PSR\, J0614$-$3329, J1231$-$1411, and J2017+0603 were discovered during radio follow-ups of unidentified {\rm Fermi} sources (Ransom et al.\ 2011; Cognard et al.\ 2011),  whereas PSR\, J0613$-$0200 was discovered in radio by Manchester et al.\  (1996), prior to its detection as a $\gamma$-ray pulsar (Abdo et al.\ 2009).

All of them are also detected in the X rays by either {\em Suzaku}, {\em Chandra}, or \xmm\ (see Abdo et al.\ 2013 and references therein). 
All these pulsars, with the exception of PSR\, J0614$-$3329, are in tight binary systems with orbital periods of $\sim$ 2 days or less and have almost circular orbits (Table \ref{orb}), as expected from the outcome of the mass-accretion phase onto the MSP.  The pulsar mass functions suggest low-mass companion stars with minimum masses $M_{\rm C}\sim0.13$--$0.3 M_{\odot}$, after assuming a pulsar mass $M_{\rm P}$=1.4$M_{\odot}$ and an orbital inclination $i=90^{\circ}$ as indicative values. These assumptions are required since for none of these MSPs it was possible to measure the post-Keplerian parameters so far and  for none of them evidence of radio eclipses has been found from the available observations (Verbiest et al.\ 2009; Cognard et al.\ 2011; Ransom et al.\ 2011).  

Deep optical investigations with 8m-class telescopes have never been reported for these four MSPs.  For both PSR\, J0613$-$0200 and PSR\, J1231$-$1411, optical observations were performed in January 2010 with the 2.4 m Isaac Newton Telescope (INT) at the La Palma Observatory (Canary Islands) soon after their detection as $\gamma$-ray pulsars  
(Collins et al.\ 2011), whereas for PSR\, J2017+0603, the only optical observations are those performed with the {\em Swift} Ultraviolet Optical Telescope  (Cognard et al.\ 2011). In all cases, no candidate companion star to the MSP was detected.

\begin{table*}
\begin{center}
\caption{Orbital parameters (orbital period P$_{\rm b}$ and eccentricity $e$) of the pulsars listed in  Table \ref{psr} as derived from the ATNF pulsar data base (second and third columns), together with the  recomputed lower limits on the companion mass ($M_{\rm C}$) inferred from the system mass function and assumed a pulsar mass $M_{\rm P}$=1.4$M_{\odot}$ and an orbital inclination $i=90^{\circ}$.  Proposed classification of the companion star are given in the following column. References for the binary system mass function measurements and  proposed classifications are given in the last column.}
\label{orb}
\begin{tabular}{lcccll} \hline
Pulsar       & P$_{\rm b}$ & $e$           & $M_{\rm C}$   & Class. & Refs. \\ 
             & (d)        &           & ($M_{\odot}$) &        & \\
\hline
J0613$-$0200 & 1.198      & 0.0000055             & 0.13        & He WD  & Verbiest et al.\ (2009)\\
J0614$-$3329 & 53.584     & 0.0001801             & 0.28        & WD  & Ransom et al.\ (2011) \\ 
J1231$-$1411 & 1.860      & 0.000004             & 0.19       & WD  & Ransom et al.\ (2011) \\ 
J2017+0603   & 2.198      & 0.0000005           & 0.18        & He WD  & Cognard et al.\ (2011) \\
\hline
\end{tabular}
\end{center}
\end{table*}


\subsection{Observation Description}

We downloaded broad-band images of the MSP fields from the public Gemini science archive\footnote{{\tt http://cadcwww.dao.nrc.ca/gsa/}}. Observations were taken between August 5 2010 and January 29 2011 with the Gemini-South telescope on Cherro Pachon (Chile) under programme GS-2010B-Q-56. Observations were performed using the Gemini Multi-Object Spectrograph (GMOS).  At the time of the observations, the camera was still mounting the original three-chip EEV CCD detector (2048$\times$4068 pixels each). This had a combined (unvignetted) field--of--view of $5\farcm5 \times 5\farcm5$, with gaps  of $2\farcs8$ between each chip. The pixel scale was 0\farcs1454 for a $2\times2$ binning. Observations were performed through the g\_G0325 ($\lambda=4750$\AA; $\Delta \lambda=1540$\AA), r\_G0326 ($\lambda=6300$\AA; $\Delta \lambda=1360$\AA), and i\_G0327 ($\lambda=7800$\AA; $\Delta \lambda=1440$\AA) filters, very similar to the g', r', and i' used by the Sloan Digitised Sky Survey (SDSS; Fukugita et al.\ 1996).  

For all targets  the observation sequences in each filter consist of both short (typically 30 s) and long (240 s) exposures, with the former taken not to saturate bright stars in the field that can be used as secondary photometric reference frames. The exposure sequence was performed applying a dithering of a few arc seconds along the X axis of the detector both to compensate for the gap between the CCD chips and account for the fringing affecting the i band. All targets were observed close to the zenith, with airmass mostly below 1.2,  in dark time and mostly in photometric conditions. The complete observation log is reported in Table 3.  

\begin{table*}
\caption{Observation log for the Gemini MSP observations}
\begin{center}
\begin{tabular}{lcclll}
\hline
PSR & Date                   & Filter               &Number of Exposures & Airmass  & seeing \\
         &  yyyy-mm-dd  &                        & $\times$ Exposure Time (s)     &       &    (\arcsec)      \\ \hline
J0613$-$0200  &	2010-11-06  &	g	 &1x30			 &	1.14   &   0.77 \\
          &        		 &	r	 &1$\times$5+1$\times$15+1$\times$30	 &1.14--1.15   & 0.62 \\
          &      		 &	i	 &1$\times$5+1$\times$30		 &	1.5  &  0.61 \\
           &      2010-12-29 &	g	 &7$\times$30+7$\times$60+7$\times$120	 &1.14--1.18  &   0.94--1.20\\
             &     			 &r	 &7$\times$30+7$\times$60+7$\times$120	 &1.13--1.14   &  0.81--1.18 \\
           &			 &		i	 &7$\times$30+7$\times$60+7$\times$120	 &1.14--1.17   & 0.63--1.00 \\ \hline
J0614$-$3329	  &	2010-09-08 &	g	 &10$\times$240			 &1.1--1.58   &  0.72--1.18\\
             &     		 &r        &10$\times$240			 &1.09--1.50   & 0.59--0.94\\
            &      		 &	i	 &8$\times$240		 &	1.12--1.45   & 0.69--0.83 \\
             &	2010-09-14 &	i	 &1$\times$30		 &		1.02    & 0.85\\
	 &			 &	i	 &2$\times$240		 &	1.17--1.19   & 0.63\\
	 &	2010-10-28 &	g	 &1$\times$30		 &		1.02    & 0.97\\
	 &			 &	r        &1$\times$30		 &		1.02    & 0.88\\ \hline
J1231$-$1411     &	2010-12-31 &	g	 &1$\times$30		 &		1.25   & 1.15 \\
	 &			 &	r	 &1$\times$30		 &		1.24  & 1.13 \\
	 &			 &	i	 &1$\times$30		 &		1.24	   & 1.03 \\
	 &	2011-01-03 &	g	 &6$\times$240		 &	1.33--1.37   &  0.87--1.05 \\
	   &			 &	r	 &1$\times$96+4$\times$240	 &	1.30--1.68    & 0.80--1.18\\
	     &    	 	 &		i &	4$\times$240	 &		1.39--1.57   &  0.76--0.81\\ 
	 &	2011-01-29 &	g	 &6$\times$240		 &	1.04--1.08   &  0.84--1.21\\
	 &			 &	r	 &6$\times$240		 &	1.04--1.1   &  0.89--1.28\\
	 &			 &	i	 &6$\times$240		 &	1.04--1.1   &  0.82--1.13 \\	     \hline
J2017+0603	  &	 2010-08-05 &	g	 &1$\times$92+1$\times$200+8$\times$240 &	1.24--1.35	  &  0.85--1.15 \\
	 &	 		 &	r	 &1$\times$87+2$\times$240		 &1.24   &  0.85\\
	 &	 2010-08-10 &	r	 &7$\times$240		 &	1.49--1.66	   &     0.85--1.05    \\	          
	   &       		 &	i	 &10$\times$240		 &	1.33--1.47   &  0.60--0.89 \\
	     &     2010-08-31 &	g &	1$\times$30		 &		1.24   &  0.95\\
	       &   		 &	r	 &1$\times$30		 &		1.24  &  0.82 \\
 	 &			 &	i	 &1$\times$30		 &		1.24  & 0.80 \\ \hline
	\end{tabular} 
\end{center}
\label{obs}
\end{table*}

\begin{figure*}
\centering
\begin{tabular}{cc}
\subfloat[J0613$-$0200]{\includegraphics[width=8cm]{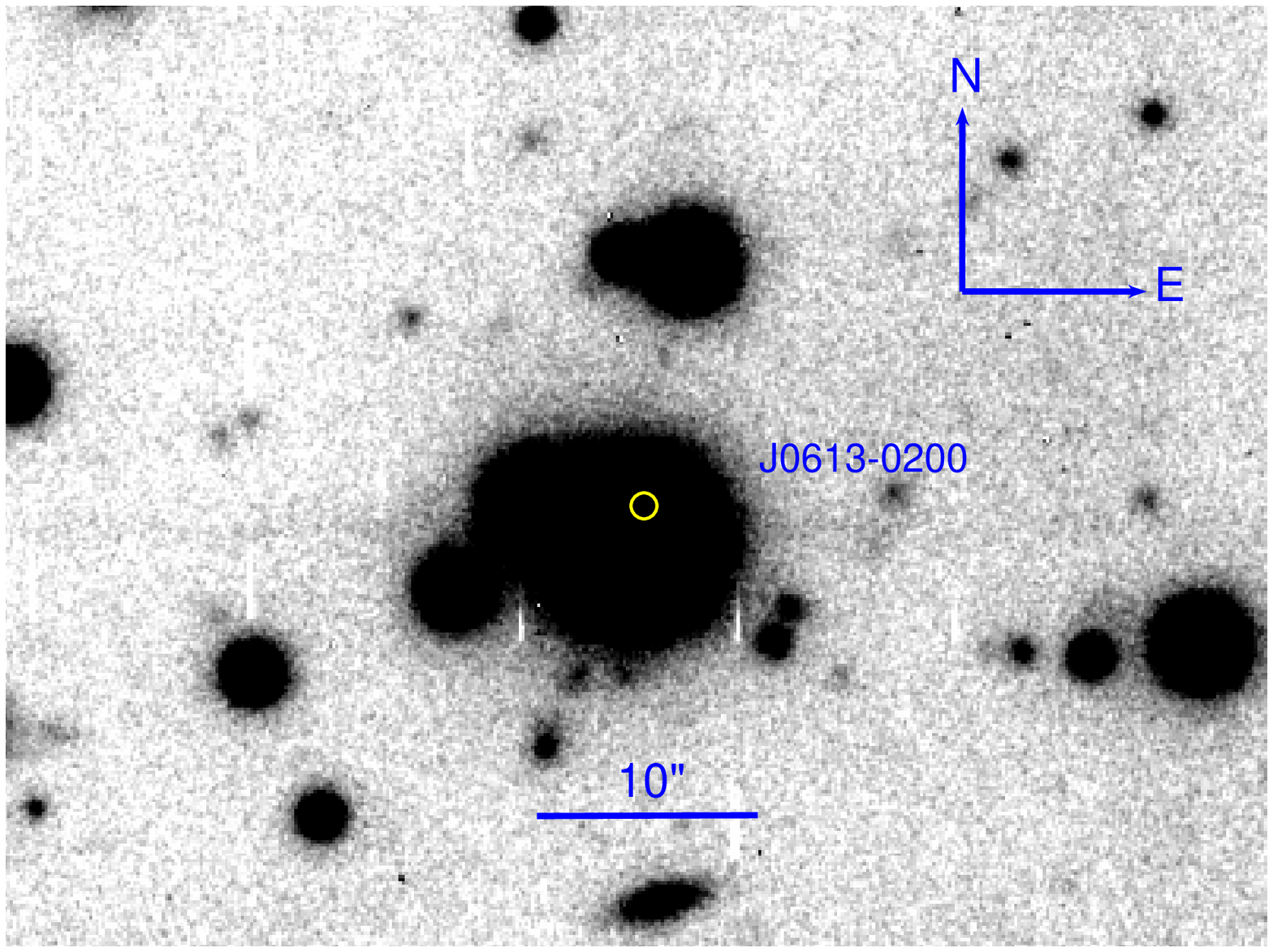}} &
\subfloat[J0614$-$3329]{\includegraphics[width=8cm]{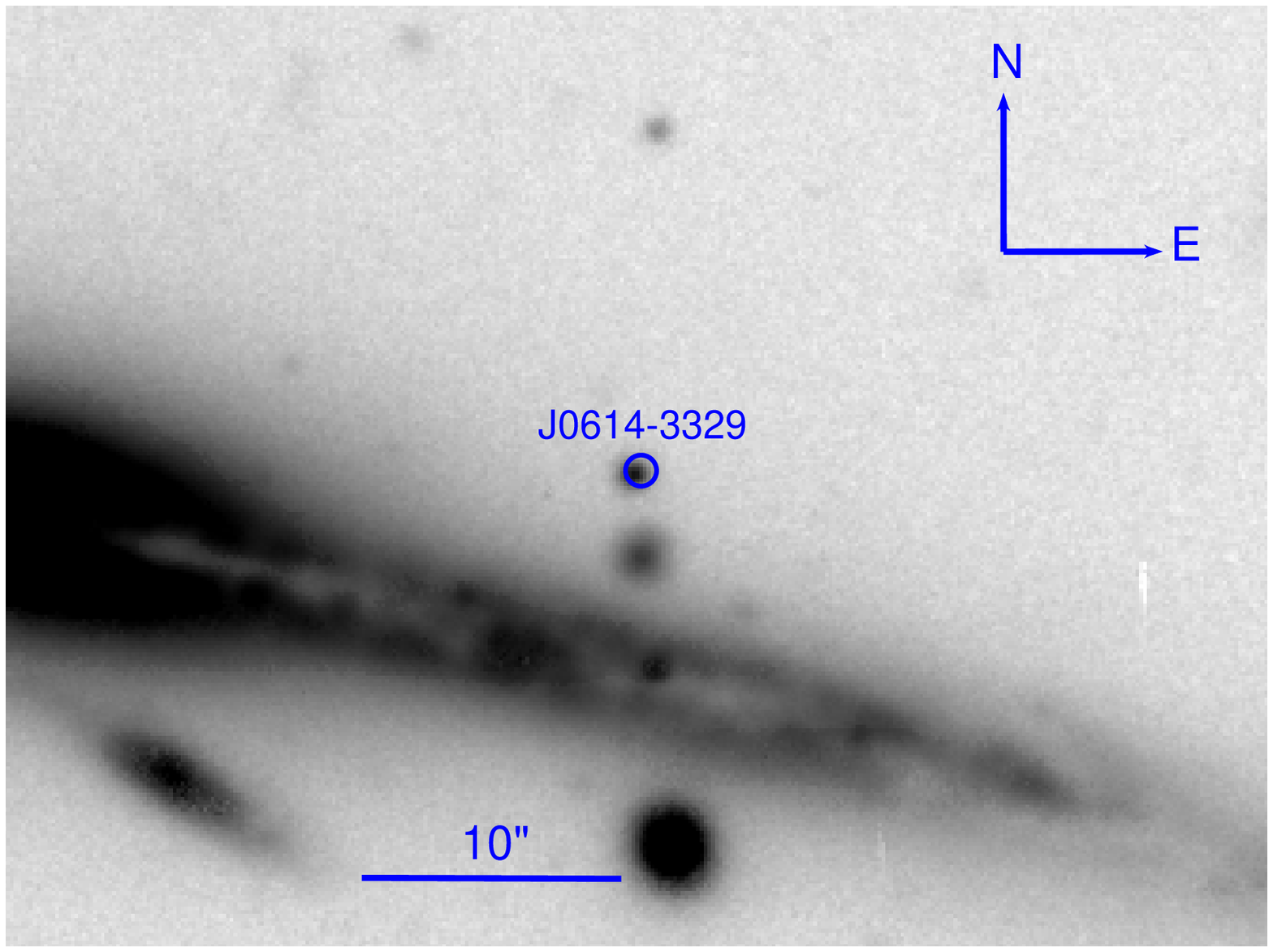}} \\
\subfloat[J1231$-$1411]{\includegraphics[width=8cm]{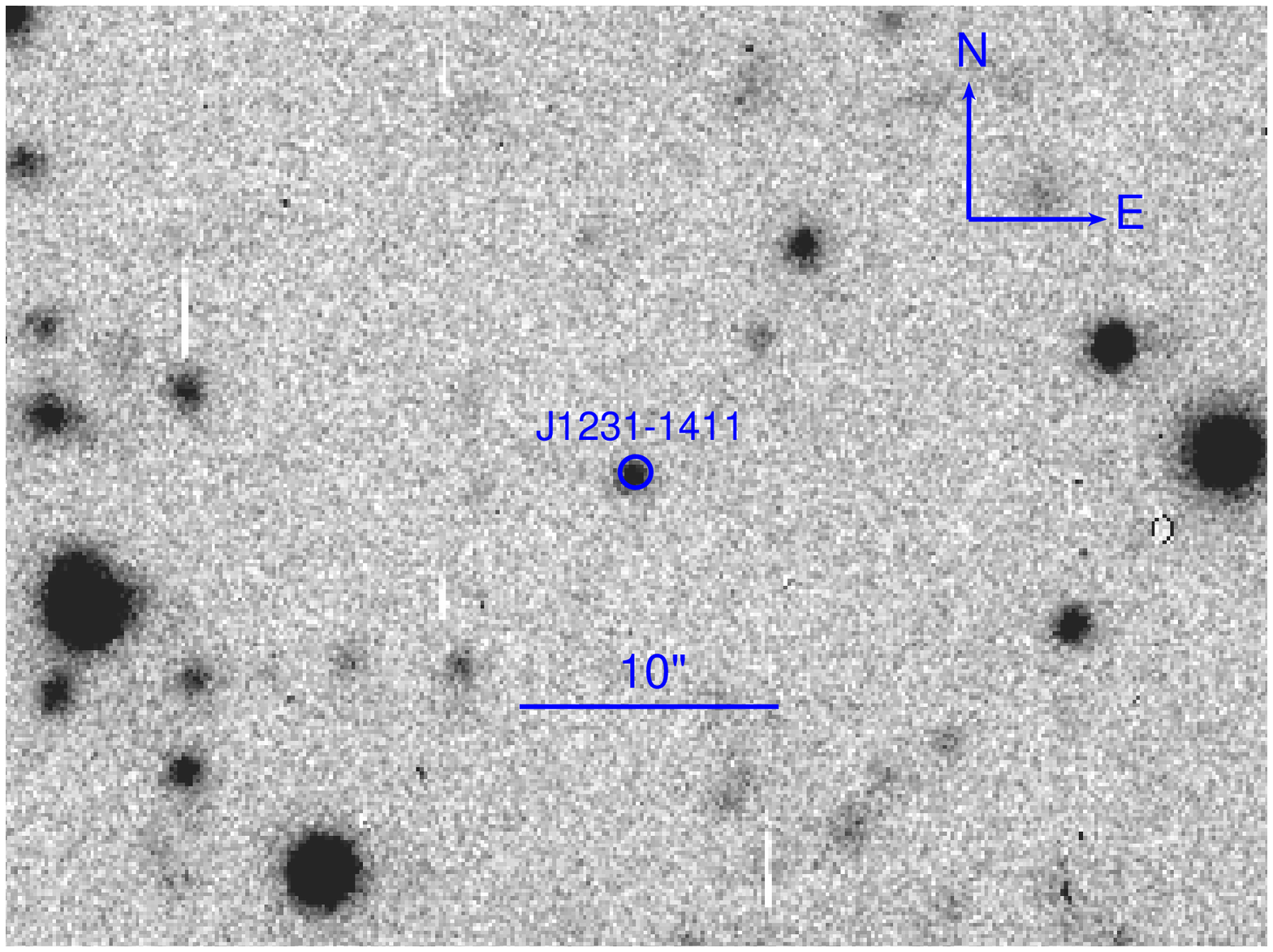}} & 
\subfloat[J2017+0603]{\includegraphics[width=8cm]{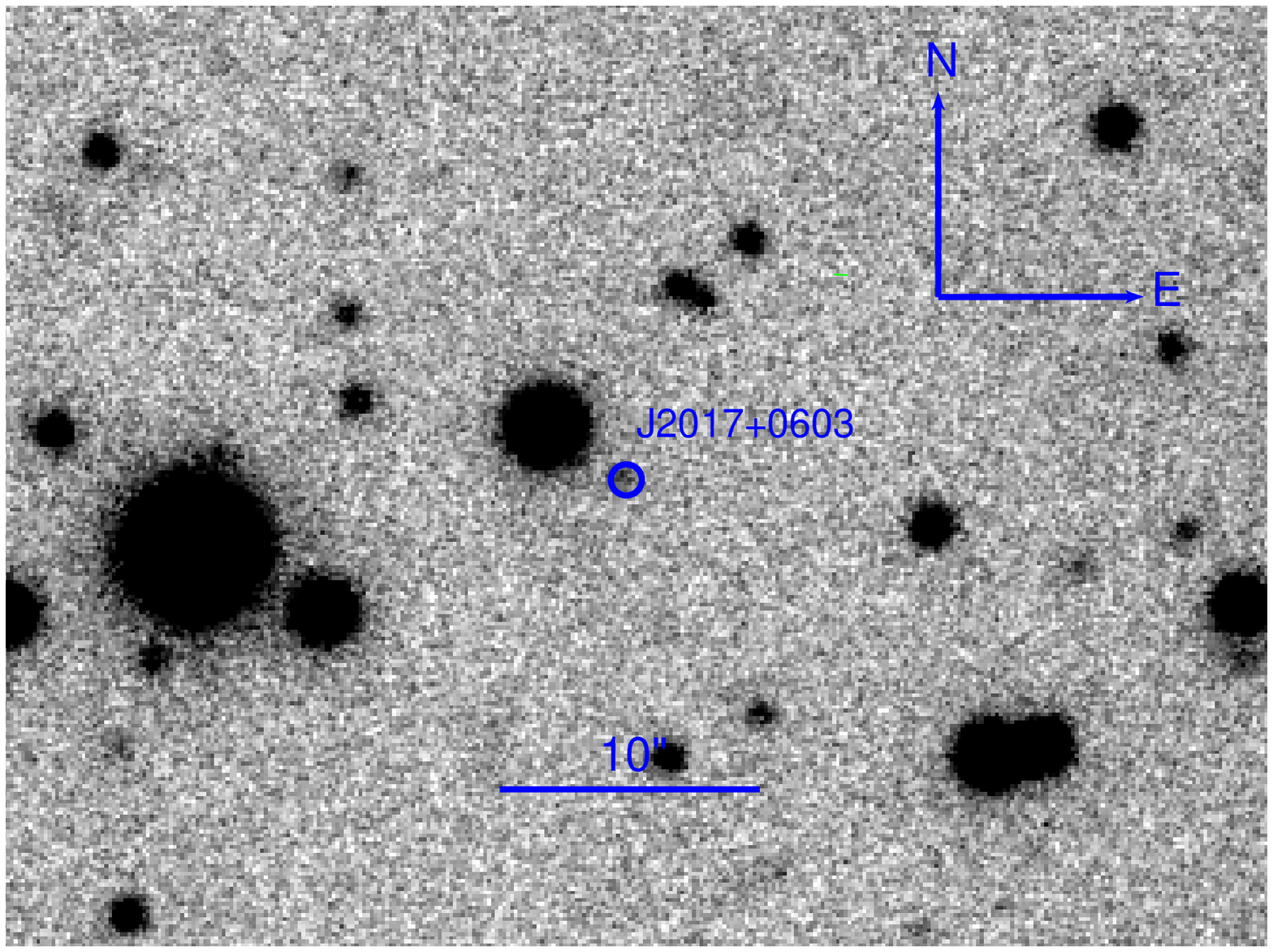}}\\
\end{tabular}
\caption{\label{all_fc} GMOS r-band images of the four MSP fields.  
The circles correspond to the computed MSP radio positions. Since the error on the radio coordinates is below 1 mas (see Table\, 1) the uncertainty on the computed radio position is dominated by the accuracy our astrometry calibration (0\farcs2; Sectn.\ 3.2). The circle radius corresponds to three times such an uncertainty for a better visualisation.
For the other pulsars, the companion stars are marked by the ticks and labelled by the pulsar names.  PSR\, J0614$-$3329 is located a few arcsec north of the disk of the bright galaxy ESO\, 365-1 (Lauberts \& Valentijn 1989).  All images have been obtained from the co-addition of all r-band science frames, apart from the  PSR\, J0613$-$0200 one which has been obtained from the co-addition of short exposures only to avoid saturation of the bright star near the pulsar position. 
}
\end{figure*}

\subsection{Data Reduction and Calibration}

We reduced the GMOS images using the dedicated  {\sc gmos}  image reduction package available in the Image Reduction and Analysis Facility ({\sc IRAF}) package\footnote{IRAF is distributed by the National Optical Astronomy Observatories, which are operated by the Association of Universities for Research in Astronomy, Inc., under cooperative agreement with the National Science Foundation.}. After  downloading the closest--in--time bias and sky flat field frames from the Gemini  science archive, we used the  tasks {\tt gbias} and {\tt giflat} to process and combine the bias and flat-field frames, respectively.  We  then reduced the single science frames using the task {\tt gireduce} for bias subtraction, overscan correction, image trimming and flat field normalisation.  These tasks work on a chip-by-chip basis and produce fully reduced images for each of the three GMOS CCDs. From the reduced science images, we produced a mosaic of the three GMOS CCDs using the task {\tt  gmosaic}. Finally, we aligned the reduced image mosaics to compensate for the dithering pattern applied to the exposure sequence and  average-stacked the aligned images with the task  {\tt imcoadd} to filter out cosmic ray hits. Unfortunately, the dithering pattern chosen for the observations happened not to be adequate to account for the effects of the fringing in the i band. For this reason, the i-band images obtained after frame stacking were still significantly affected by strong fringing. Applying the stacking to different image sub-groups did not improve  the results significantly. To remove the fringing, we followed the recipe suggested by the GMOS science operation team\footnote{{\tt http://www.gemini.edu/sciops/instruments/gmos/calibration/\\example-cal-data?q=node/10456}}. We downloaded from the Gemini science archive
the template i-band fringe images closest in time to the epochs of our observations. Then, after flux normalisation, we subtracted the fringe images from the single science images using the {\sc IRAF} task {\tt girmfringe}. Finally, we applied the same procedures as above to mosaic and stack the de-fringed i-band science images.

We computed the astrometry calibration of the optical images (both image mosaics and stacks) using the {\em wcstools}\footnote{{\tt http://tdc-www.harvard.edu/wcstools/}} suite of programs that automatically match the sky coordinates of stars selected from, e.g the Two Micron All Sky Survey (2MASS; Skrutskie et al.\ 2006) or the Guide Star Catalogue 2.3 (GSC2; Lasker et al.\ 2008) catalogues with their pixel coordinates computed by {\em Sextractor} (Bertin \& Arnouts 1996).    After iterating the matching process and applying a sigma-clipping selection to filter out obvious mismatches, high-proper motion stars, and false detections, a pixel--to--sky coordinate transformation was computed using a  second order polynomial function and we obtained,  for the ground-based images mean residuals of $\sim 0\farcs2$ in the radial direction, using 50 bright, but non-saturated, 2MASS stars. To this value we added in quadrature the uncertainty $\sigma_{tr}$ = 0\farcs08 of the image registration  on the  2MASS reference frame. This is given by $\sigma_{tr}$=$\sqrt{n/N_{S}}\sigma_{\rm S}$ (e.g., Lattanzi et al.\ 1997), where $N_{S}$ is the number of stars used to compute the astrometric solution, $n$=5 is the number of free parameters in the sky--to--image transformation model, $\sigma_{\rm S} \sim 0\farcs2$ is the mean absolute position error of  2MASS  for stars in the magnitude range  $15.5 \le K \le 13$ (Skrutskie et al.\ 2006). After accounting for the 0\farcs015  uncertainty on the link of 2MASS to the International Celestial Reference Frame  (Skrutskie et al.\ 2006), we ended up with an overall accuracy of $\sim$0\farcs2 on the absolute optical astrometry, i. e. comparable to the GMOS pixel size. Given the exquisite accuracy of MSP radio coordinates (Table \ref{psr}) the uncertainty on the absolute astrometry calibration of the GMOS images obviously dominates the accuracy on the positional coincidence with a potential companion star.

We computed the photometry calibration using as a reference sets of secondary photometric standard stars extracted from the  American Association of Variable Stars Observers (AAVSO) Photometric All-Sky Survey\footnote{{\tt http://www.aavso.org/apass}} (APASS), directly identified in each GMOS image of our MSP fields. We preferred to follow this approach rather than computing the photometry calibration using images of standard star fields (Smith et al.\  2007) because there were no standard star observations taken on the same nights as the science observations. Moreover, a direct on--the--frame calibration obviously allows one to compensate for variations in the sky transparency during the night.
The APASS photometric survey (Henden \& Munari 2014) is calibrated on the SDSS photometric system and is, then, well suited to calibrate our GMOS g, r, i-band images. Like the SDSS, our photometry is, then, in the AB system (Oke 1974).

\section{Results}

\subsection{Astrometry}

By using the astrometric solutions obtained as described in the section above, we checked all co-added science frames in each filter to look for objects detected at the computed radio position of our target MSPs (Table 1), which can be be regarded as potential  companion stars. For those pulsars which have a measured proper motion, i.e. PSR\, J0613$-$0200  (Verbiest et al.\ 2009) and PSR\, J1231$-$1411 (Ransom et al.\ 2011), we corrected the reference radio positions to the epoch of the Gemini observations.

For three pulsars  (PSR\, J0614$-$3329, J1231$-$1411, J2017+0603), we clearly detected an object at the radio coordinates (see Fig.\ 1; top right to bottom left). In all cases the match was very good, with the object centroid almost perfectly coincident with the computed pulsar radio position. In particular, the measured angular separation between the MSP coordinates and those of its candidate companion was only $\sim 0\farcs2$--0\farcs3, i.e. comparable with the astrometric uncertainty of the GMOS images ($\sim$ 0\farcs2; Section 2.3). Thus, in each case we considered the association with the MSP safe. 
%
To formally quantify the goodness of the associations between the MSPs and their candidate companions, we computed the probability that they are due to a chance coincidence. We computed this probability as $P=1-\exp(-\pi\rho r^2)$, where  $r$ is  the measured angular separation between the MSP coordinates and those of its candidate companion (in arc second), and $\rho$ is  the  density of  stellar  objects around the MSP positions per square arc second. Since, in principle, we do not know a priori the expected brightness range for the companion stars, we computed $\rho$ without applying a selection in magnitude.
For PSR\, J0614$-$3330 we computed a stellar density $\rho\sim0.028$  and estimated a chance coincidence probability $P\sim0.0035$, whereas for PSR\, J1231$-$1411 and PSR\, J2017+0603 we estimated $P\sim0.004$ ($\rho\sim 0.014$) and  $P\sim0.010$ ($\rho\sim 0.038$), respectively. In all cases, the low chance coincidence probability would rule out spurious associations and would suggest that the candidate companions detected in the GMOS images are indeed associated with the MSPs. 


For the fourth pulsar (PSR\, J0613$-$0200), the radio position fell slightly off-centre ($\sim 2\arcsec$ to the North) with respect to the centroid of the closest object  detected in the GMOS images (Fig.\ 1; top left), although it is still within its PSF. However, the measured position offset is much larger than the $3 \sigma$ uncertainty on our computed astrometric solution, with the uncertainty on the pulsar radio coordinates being negligible (Table 1). Thus, based on our astrometric solution, we infer that this object is not associated with  PSR\, J0613$-$0200.  Presumably, this object is the "brightish star nearby" the pulsar, mentioned in Table\, 1 of  van Kerkwijk et al.\  (2005). This star was also identified by Mignani et al.\ (2014) in near-ultraviolet and optical images from the Galaxy Evolution Explorer ({\em GALEX}), the  \xmm\ Optical Monitor (OM), and the  {\em Swift}  UltraViolet and Optical Telescope (UVOT) and ruled out as a candidate companion to the pulsar on the basis of its positional offset from the radio coordinates.  The same star was also detected in images taken with the INT (Collins et al.\ 2011) but was considered unassociated to PSR\, J0613$-$0200 for the very same reason as above.   The relatively high chance coincidence probability 
with the radio position also suggests that this star is unrelated to the pulsar. By assuming a radius $r\sim$ 2\arcsec, i.e. equal to the angular separation between the star and the MSP radio coordinates, we obtain $P\sim 0.13$ ($\rho \sim 0.038$). 
To directly verify a possible, though unlikely, association between this star and the pulsar, we compared their measured proper motions. According to UCAC-4 (Zacharias et al.\ 2013), this star has a proper motion of  $\mu_{\alpha} cos(\delta) =7.2\pm2.8$ mas yr$^{-1}$ and $\mu_{\delta}=-4.6\pm2.9$ mas yr$^{-1}$ in right ascension and declination, respectively, which is marginally consistent with the pulsar radio proper motion, $\mu_{\alpha} cos(\delta) =1.84\pm0.08$ mas yr$^{-1}$ and $\mu_{\delta}=-10.6\pm0.2$ mas yr$^{-1}$ (Verbiest et al.\ 2009). However, the UCAC-4 
measurement is still below the $3 \sigma$ significance so that the result of such a comparison is not conclusive. 
Thus, with all pieces of evidence pointing towards a chance coincidence, we conclude that the pulsar companion
is most likely hidden in the PSF wings of this bright star.
%


\subsection{Photometry}

Since for all the target MSPs, the radio position is neatly located very close to the nominal GMOS pointing position at the centre of chip \#2, we decided to use only images from the central chip for the following photometry analysis. This choice has the advantage of  simplifying the procedure, with the tolerable cost of loosing the relatively small field of view of the two external chips, some of which are also affected by vignetting and partially occulted by the telescope auto-guider. Moreover, adopting a smaller, but still significantly large field of view, has the advantage of reducing the effects of possible differential extinction along the line of sight due to, e. g. clumps in the interstellar medium.

We computed the photometry of the candidate MSP companion stars,  as well as of the field stars, through a semi-automatic pipeline that makes use of the {\textsc DAOPHOT II} package (Stetson 1987, 1994) distributed as part of {\sc IRAF} software
environment. The pipeline consists of a set of scripts executing sequentially all the tasks needed to obtain a list of objects from a given image, with positions, magnitudes and errors. Namely, per each filter these tasks are:
i) obtain a master list of detected objects through the analysis of the master image, chosen as the co-added and exposure-map corrected science frame  to maximise the detection signal--to--noise, ii) obtain a reliable PSF model for all the images in the data set (single exposures and co-added science frames), iii) perform the PSF-fitting photometry and aperture correction steps on both single exposures and co-added science frames to obtain magnitudes and errors of every object in the master list, iv) apply the photometric calibration as described in the previous Section, v) extract relevant data for the candidate MSP companion stars and some control objects selected among the closest neighbour stars, and, per each MSP field,  vi) perform a variability analysis of the different photometry measurements to spot possible flux variations of the candidate MSP companion stars. 
The only step of our photometry pipeline which is run manually is the selection of the objects for building the PSF model in step ii), from the master list obtained from the master image. Once the starting sample has been selected, the actual PSF modelling is automatised. This procedure is a "custom" version of what more specialised programs do (see e.g. {\textsc ALLFRAME}; Stetson 1994) and has been repeatedly tested in the past for similar works.

The final result of our photometry pipeline is a catalogue of positions (detector and celestial coordinates) and  g, r, i flux measurements and errors for each MSP field and for each image  (single exposures and co-added science frames) in our sample. Filter-to-filter matching was done in a trivial way because all the single-mosaic images are carefully registered onto the master. From this data set, we extracted time series for each candidate MSP companion  (Section 3.3), identified from the astrometric matching with the radio positions, and built both colour--magnitude (CM) and colour--colour (CC) diagrams using mean magnitudes for both the candidate MSP companion stars and all stars detected around the MSP positions  (Section 3.4).

Using the same approach as above, we computed a model PSF for all the PSR\, J0613$-$0200 images (single mosaics and co-added science frames). Then, we used it to subtract the star detected $\sim$2\arcsec\ from the pulsar radio position and search for possible objects hidden in its PSF wings. Unfortunately, the low signal--to--noise of the residuals after the subtraction of the star PSF did not allow us to find evidence of any excess of signal that could be associated with an object detected at the pulsar position.  Thus, the actual companion to PSR\, J0613$-$0200 would be undetected in the GMOS images. Only higher spatial resolution observations, either with Adaptive Optics or with the {\em Hubble Space Telescope}, would make its detection possible.  After subtracting the star PSF we estimated the upper limits on the flux of the PSR\, J0613$-$0200 companion star using the co-added science images as a reference. The $3 \sigma$ upper limits computed at the pulsar radio position are $g=25.0$, $r= 24.3$, and $i= 24.2$.



The companion stars to MSPs often feature optical variability along the orbital period of the binary system owing, for instance to tidal distortion of the star (e.g. Orosz \& van Kerkwijk 2003) or the formation of hot spots on the star surface  irradiated by the pulsar (e.g., Stappers et al.\ 2001). For this reason, 
we searched  
for possible evidence of variability 
in  the light curves of the candidate companion stars to PSR\, J0614$-$3329, J1231$-$1411, J2017+0603, obtained from the multi-epoch GMOS photometry (Table \ref{obs}).
 As a criterion, we used only the photometric measurements obtained in those filters where the candidate companion star is brighter and is detected with the highest occurrence. For all MSP fields, these conditions are matched by the observations in the r filter. However, in all cases  the reduced $\chi_{r}^2$ of the flux measurement distribution is quite low, with values of 0.18 (J0614$-$3330),  0.24 (J1231$-$1411),  0.06 (J2017+0603), and is compatible with random variations only.
 %
%
%
Moreover, the small number of flux measurements, concentrated in short night fractions and repeated on two or three nights only, sometimes separated by up to a few weeks (see Table \ref{obs}), largely under sample the known orbital periods.  This is particularly true in the case of PSR\, J0614$-$3329 which has the  longest orbital period (P$_{\rm b}$=53.584 d) among these three MSPs.
%
Therefore, we cannot use the available multi-epoch flux information as a  further piece of evidence to support the proposed identification between the MSPs and their candidate companion stars, so far only based on the very good positional coincidence with the pulsar radio coordinates. 

\begin{table*}
\centering
\caption{Coordinates and multi-band photometry for the candidate companions to the three MSPs (column one), as measured in the GMOS images. Magnitude values were computed from the mean of all the available measurements and are in the AB system (Oke 1974). Column four gives the angular separation $\Delta~r$ between the measured candidate companion and the MSP coordinates (Table \ref{psr}).} 
\label{stars}
\begin{tabular}{lcccccc} \hline\hline
PSR & $\alpha$  & $\delta$ & $\Delta~r$ & g & r & i  \\  
        & $^{(hms)}$  & $(^{\circ} ~ \arcmin ~ \arcsec)$ &  &  \\ \hline 
	J0614$-$3329 & 06 14 10.333 & $-$33 29 54.19 &  0\farcs19 & 21.95$\pm$0.05 & 21.70$\pm$0.03 & 21.58$\pm$0.03 \\
	J1231$-$1411 & 12 31 11.299 & $-$14 11 43.39  & 0\farcs31   &25.40$\pm$0.23  & 23.95$\pm$0.06 & 23.35$\pm$0.11 \\ 
	J2017+0603   &  20 17 22.714 & +06 03 05.82    & 0\farcs29&  24.72$\pm$0.28 & 24.06$\pm$0.25 & 23.84$\pm$0.17 \\ \hline
\end{tabular}
\vspace{0.5cm}
\end{table*}

\begin{figure*}
\centering
{\includegraphics[height=8cm,angle=0,clip=]{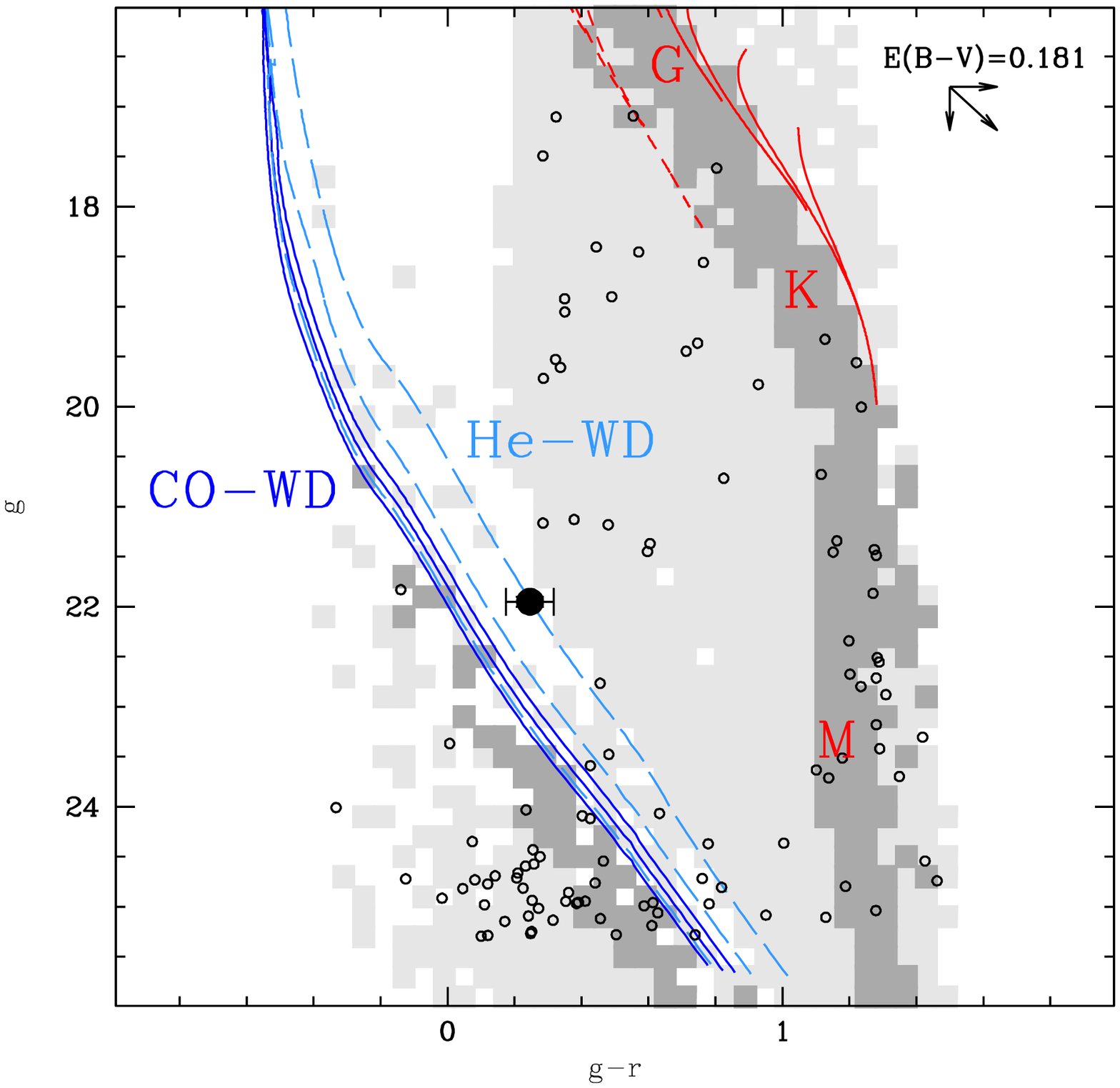}}
{\includegraphics[height=8cm,angle=0,clip=]{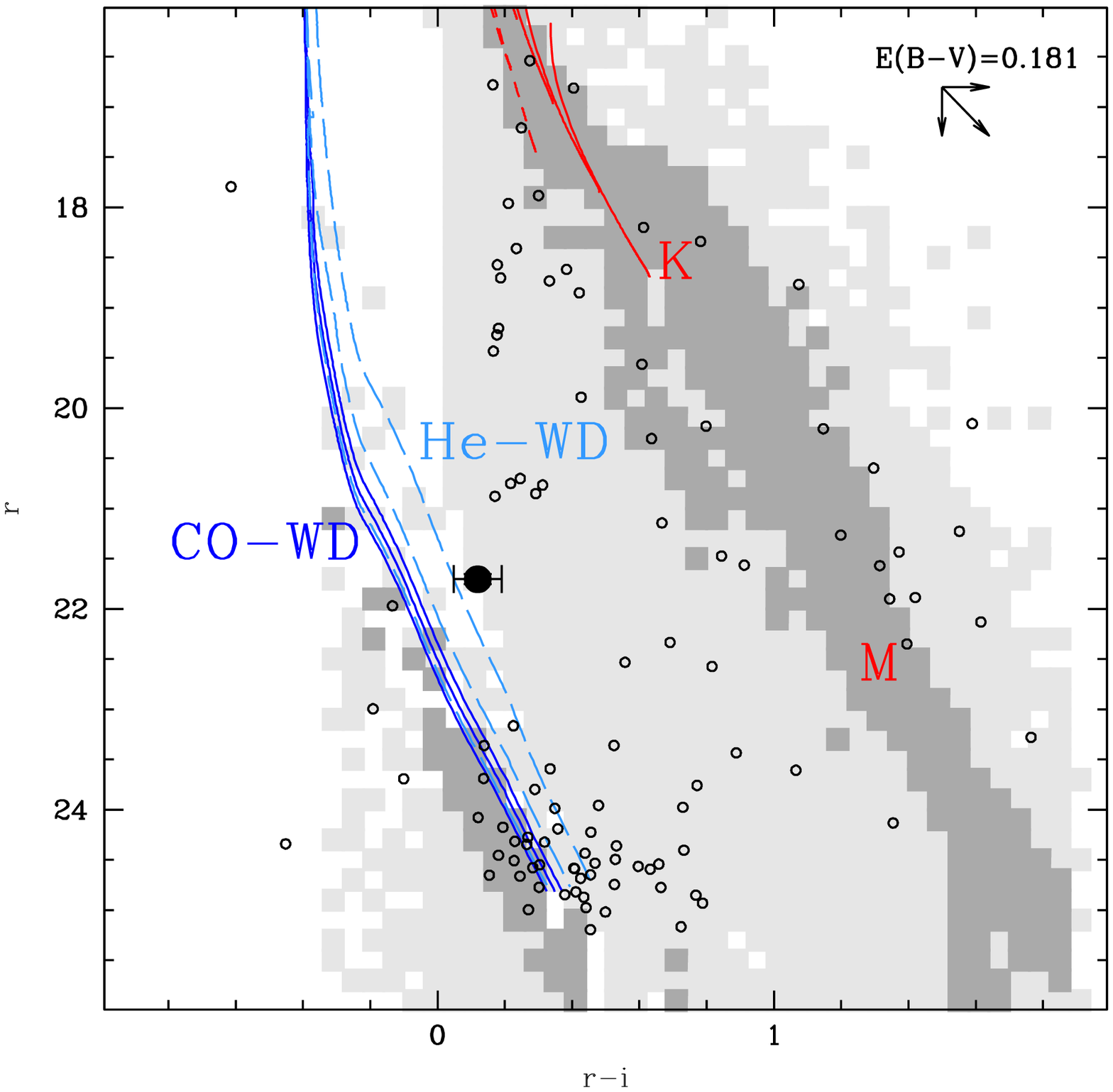}}
{\includegraphics[height=8cm,angle=0,clip=]{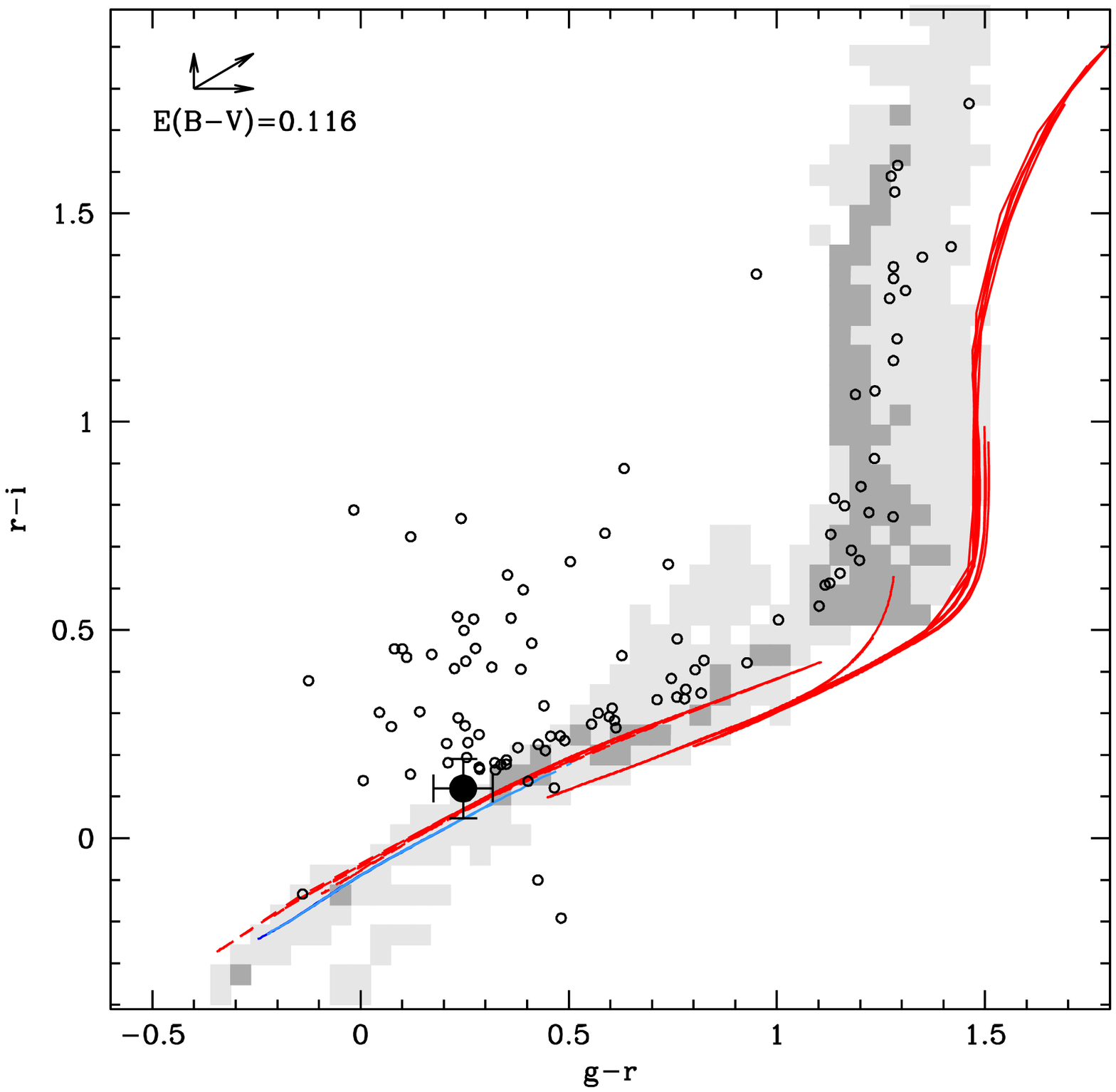}}

\caption{Observed CM (top row) and CC (bottom row) diagrams for all the stars in a $\sim 2\farcm0 \times 5\farcm5$ sky region around the PSR\, J0614$-$3329 position  (filled black circles). The candidate companion star to the pulsar  is marked by the black filled circle. Stellar sequences simulated from the Besan\c{c}on models for different values of distance are shown in light and dark grey. In the CM diagrams the dark grey regions correspond to distance values within $\pm 20\%$  the assumed pulsar distance  (see Table \ref{dist}), whereas in the CC diagram they correspond to magnitudes within $\pm$ 0.05 the r-band magnitude of the pulsar candidate companion star. 
Theoretical evolutionary tracks for both He WDs ($0.2, 0.25, 0.35 M_{\odot}$; light blue dashed lines) and low-mass CO WDs ($0.35, 0.4, 0.45 M_{\odot}$; dark blue solid lines)  are represented, computed from the models of  Panei et al.\ (2007) for an age up to 4.8 Gyr and
for the nominal values of the MSP distance. Masses increase from right to left.  Evolutionary tracks for MS stars for both Z=0.02 (red solid lines) and  Z=0.0001 (red dashed lines) are also shown for different mass ranges  (0.5--1 $M_{\odot}$, with steps of  0.1 $M_{\odot}$) and for the nominal MSP distance. Masses decrease from top to bottom. 
Only in this case we assumed a distance value half of that obtained from the DM (see discussion in Ransom et al.\ 2011).  Both the simulated stellar sequences from the  Besan\c{c}on models and the WD and MS evolutionary tracks were plotted assuming a null reddening. Reddening vectors are shown on the top of each panel. The lengths of the vectors correspond to the E(B-V) estimated  from the hydrogen column density inferred from the fits to the MSP X-ray spectrum. We used the extinction coefficients of Fitzpatrick (1999).  
  \label{0614-cmd} 
}
\end{figure*}

\begin{figure*}
\centering
{\includegraphics[height=8cm,angle=0,clip=]{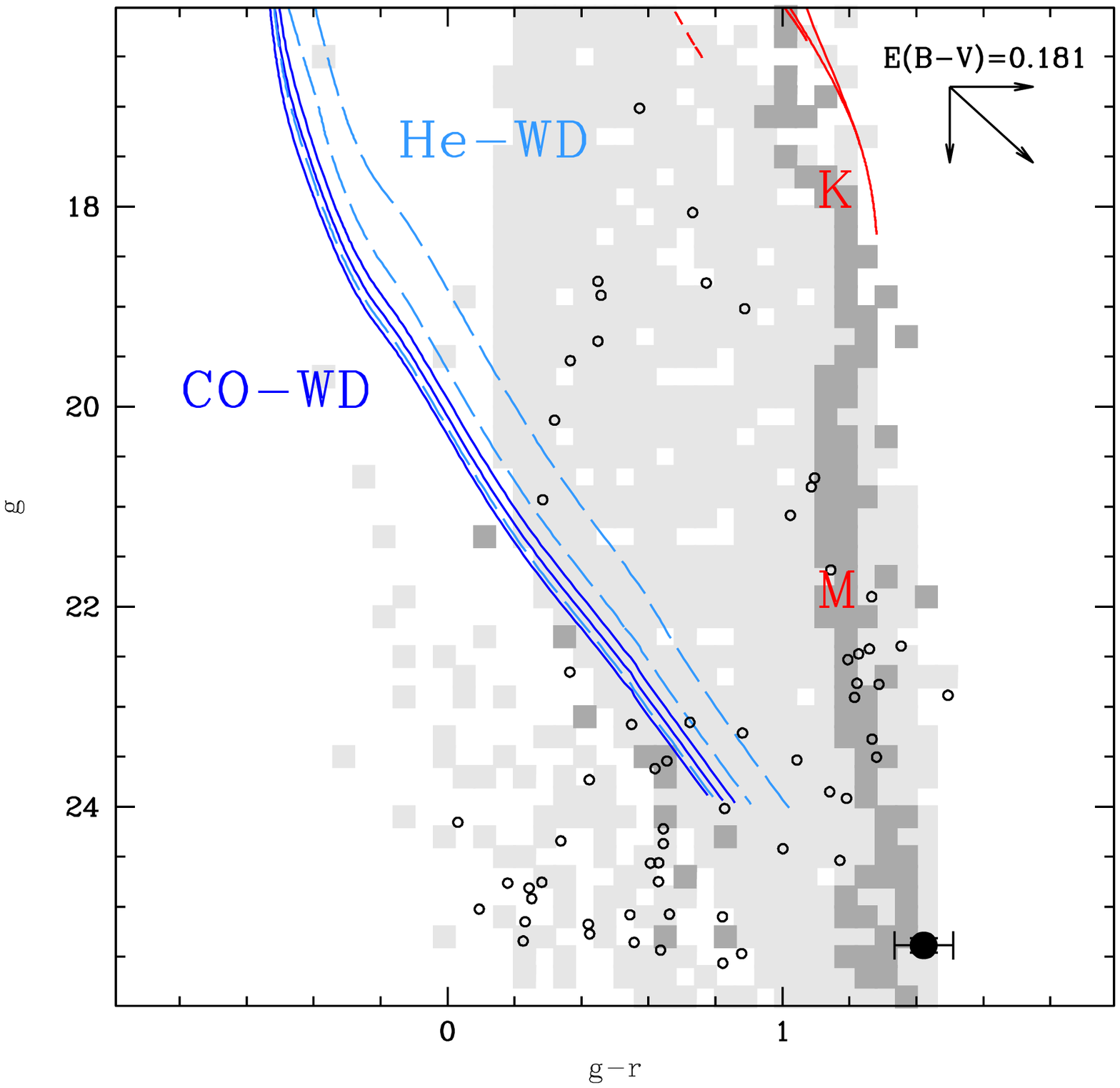}}
{\includegraphics[height=8cm,angle=0,clip=]{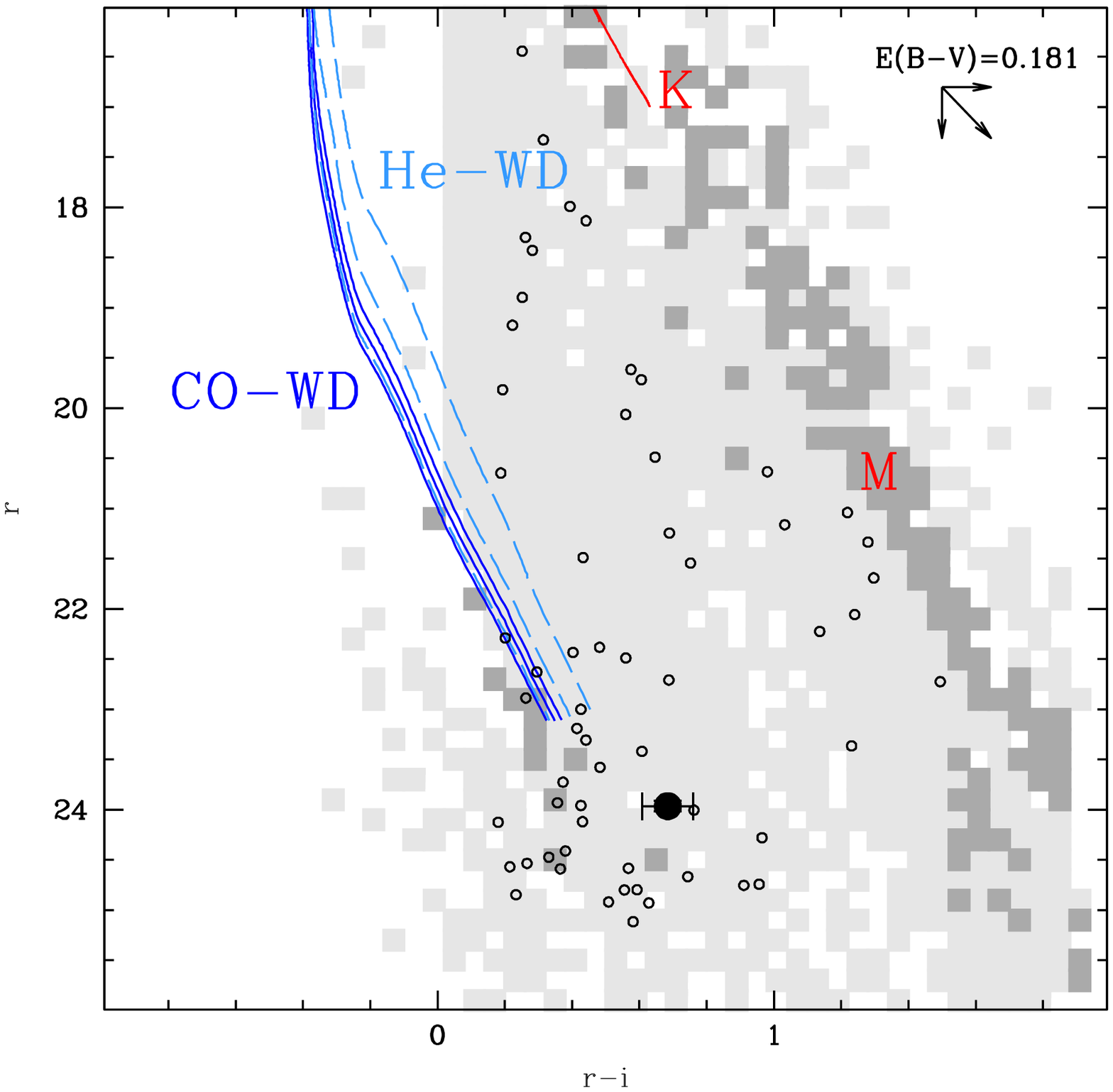}}
{\includegraphics[height=8cm,angle=0,clip=]{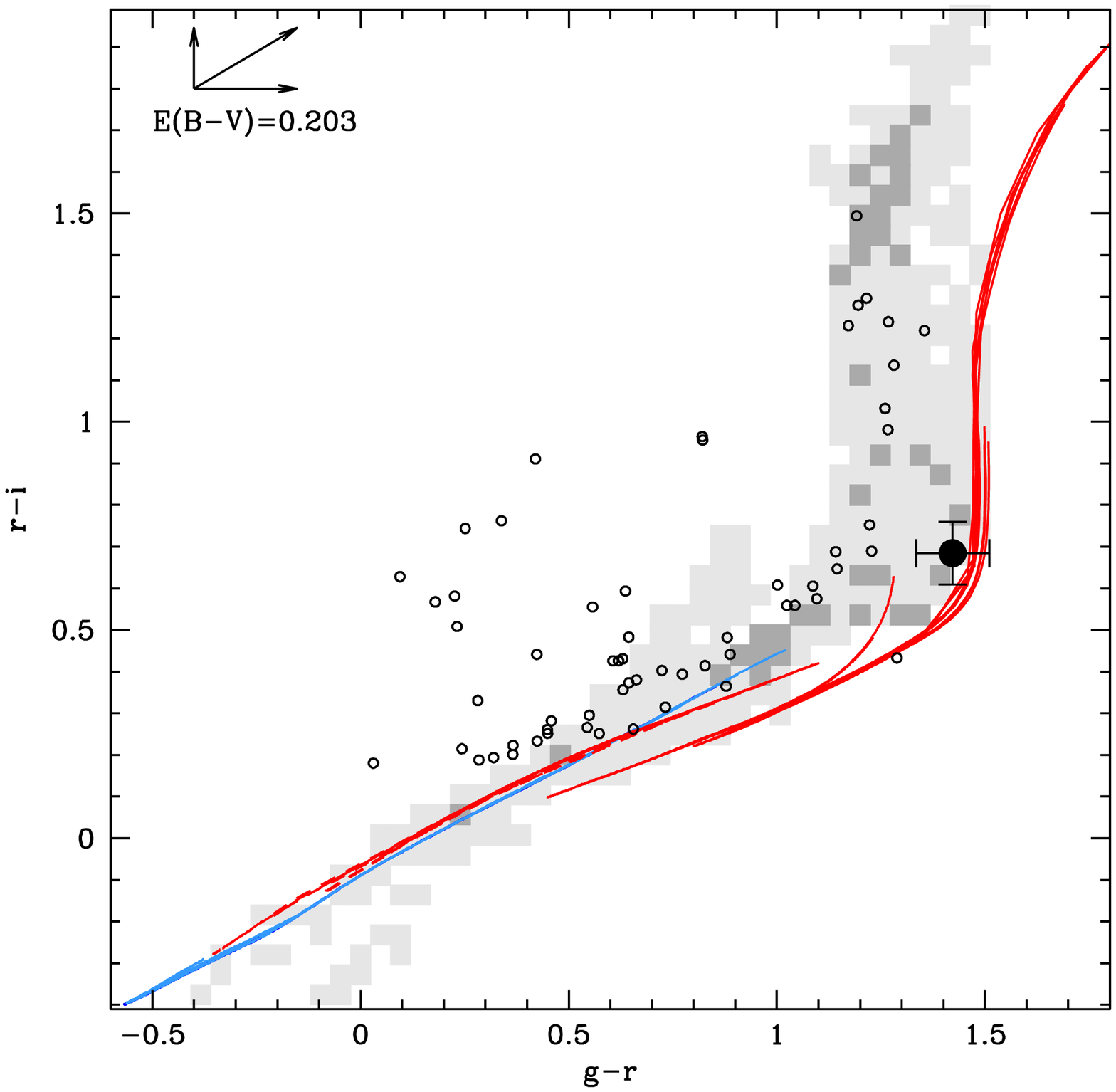}}

\caption{Same as Figure \ref{0614-cmd} but for the PSR\, J1231$-$1411 field. \label{1231-cmd} 
}
\end{figure*}

\begin{figure*}
\centering
{\includegraphics[height=8cm,angle=0,clip=]{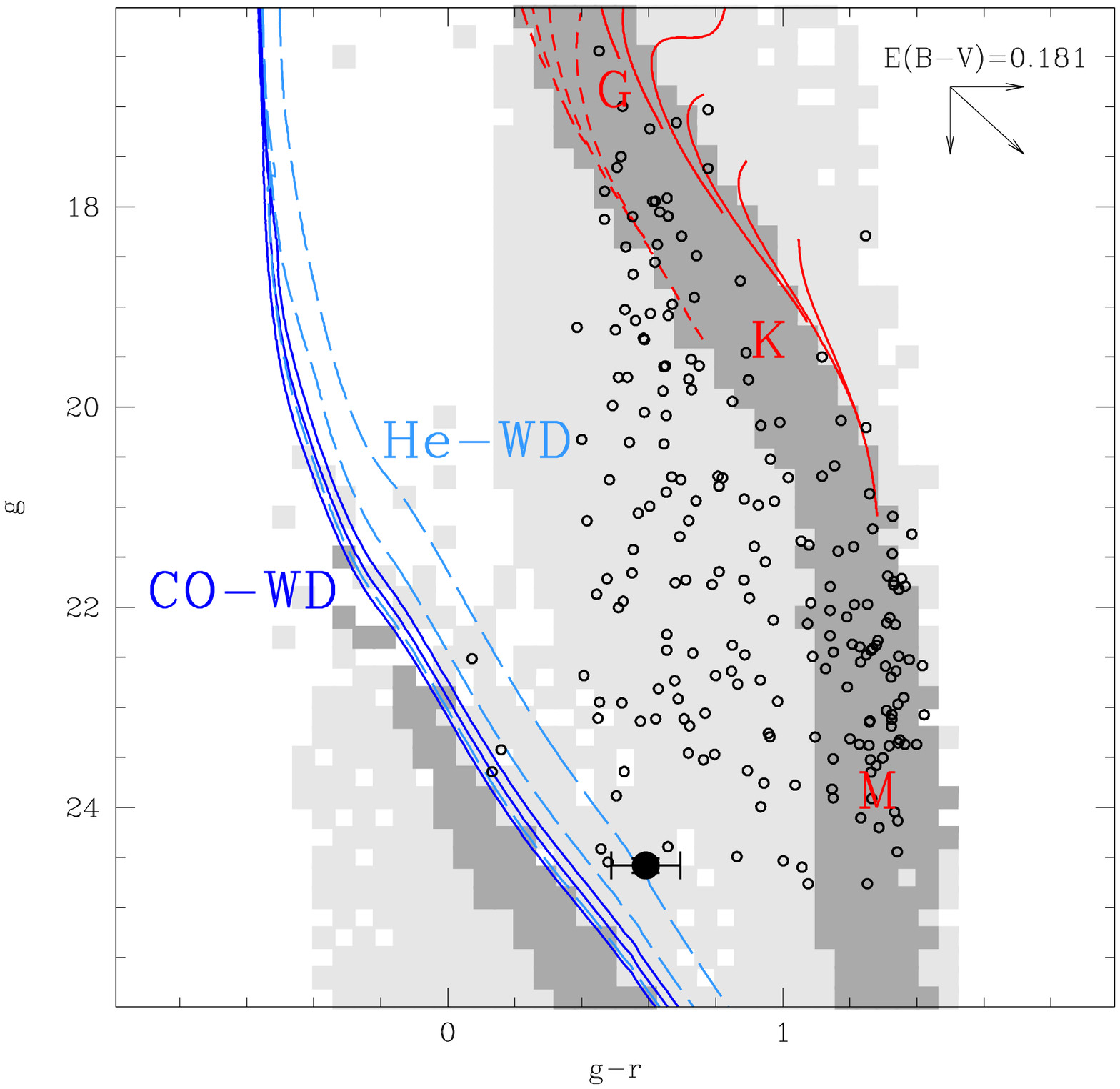}}
{\includegraphics[height=8cm,angle=0,clip=]{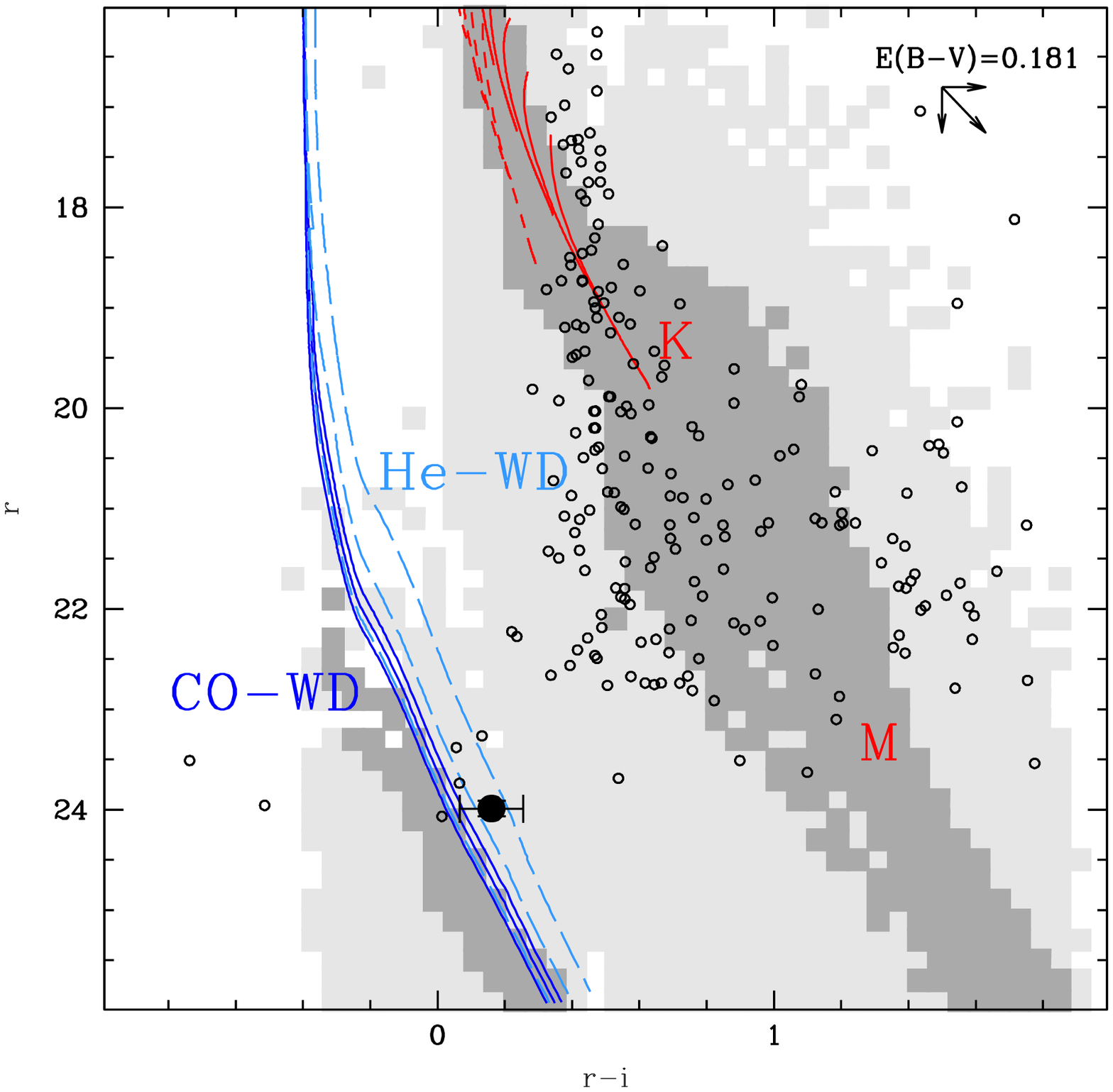}}
{\includegraphics[height=8cm,angle=0,clip=]{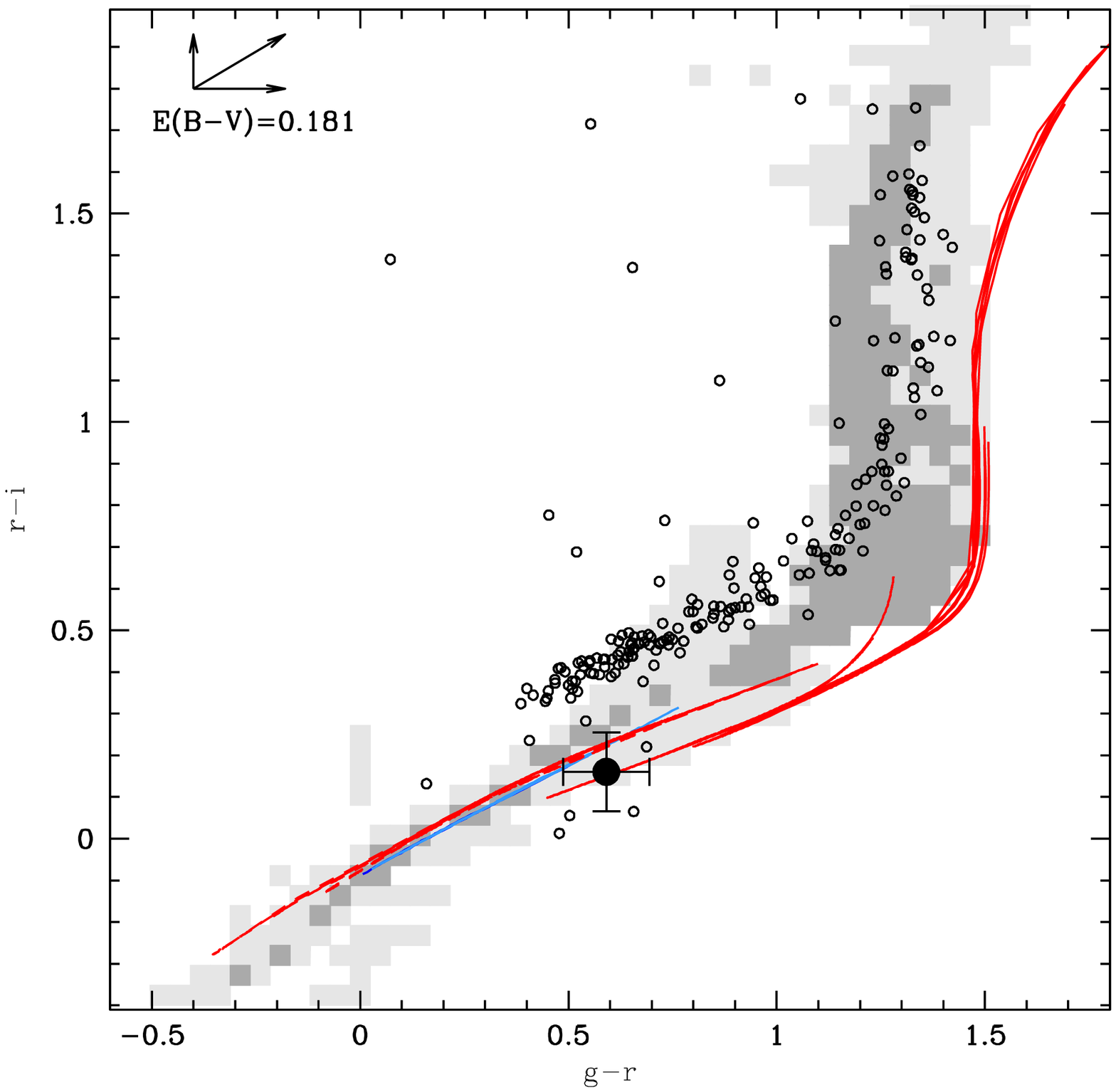}}

\caption{Same as Figure \ref{0614-cmd} but for the PSR\, J2017+0603 field.\label{2017-cmd} 
}
\end{figure*}

\subsection{Colour and magnitude analysis}

Since no significant variations are seen in the objects' fluxes, we computed mean optical magnitudes in the g, r, and i filters for the candidate companion stars to PSR\, J0614$-$3329, J1231$-$1411, J2017+0603, after applying a $\sigma$ clipping algorithm to filter out measurements more strongly affected by night--to--night fluctuations.  The mean magnitudes of the MSP candidate companion stars are (Table \ref{stars}): g=21.95$\pm$0.05, r=21.70$\pm$0.03, i=21.58$\pm$0.03 (PSR\, J0614$-$3330),  g=25.40$\pm$0.23, r=23.95$\pm$0.06, i=23.35$\pm$0.11 (PSR\, J1231$-$1411), and g=24.72$\pm$0.28, r=24.06$\pm$0.25, i=23.84$\pm$0.17 (PSR\, J2017+0603).

Incidentally, we note that  the  mean magnitudes of the star detected $\sim 2$\arcsec\ from the computed radio position of PSR\, J0613$-$0200 (Fig.\ref{all_fc}, top left) are: g=16.21$\pm$0.03, r=15.54$\pm$0.03, i=15.35$\pm$0.03. This suggests that, at the pulsar parallactic distance of 1.25$\pm$1.09 kpc
(Verbiest et al.\ 2009), it would be an early main sequence (MS) star,
%
whereas the minimum companion star mass and the orbital parameters
(Table \ref{orb}) 
indicate a WD companion star to  PSR\, J0613$-$0200, possibly an He WD (Lorimer et al.\ 1995).  Moreover, MSPs with possible early MS companions  are extremely rare, with  PSR\, J1903+0327 being the only certified case so far (Freire et al.\ 2011). However, this MSP has a quite large orbital eccentricity  ($e\sim0.44$) and a long orbital period ($P_{\rm b}$= 95.17 d), whereas  PSR\, J0613$-$0200 is in an almost circular orbit with a short period (Table \ref{orb}). This indicates  that the two binary systems followed different evolutionary paths, hence with different companions, with PSR\, J1903+0327 probably being part of a triple system in origin with both the MS star and a WD  (Freire et al.\ 2011). Therefore, the colour and magnitude analysis 
give a further piece of evidence that this star is unrelated to PSR\, J0613$-$0200.
An updated optical proper motion measurement for this star (see Sectn.\, 3.1), together with radial velocity measurements from optical spectroscopy and/or multi-epoch optical photometry, would  indisputably rule out its association with PSR\, J0613$-$0200.


For the other pulsars (PSR\, J0614$-$3329, J1231$-$1411, J2017+0603) we used the mean magnitudes of the candidate companion stars  as a reference for their classification by analysing their locations in the observed (i.e. not corrected for the reddening) CM and CC diagrams.
The observed CM and CC diagrams for  the PSR\, J0614$-$3329, J1231$-$1411, and J2017+0603 fields are shown in Figure \ref{0614-cmd}--\ref{2017-cmd} (black filled circles). The locations of the candidate MSP companion stars are shown as blue filled circles with error bars. In order to reject outliers and include only high-confidence measurements, we plotted stars for which at least nine measurements per filter are available and with $\sigma < 0.08$.

As seen, also owing to the limited star sample, all the observed CM diagrams show a significant scatter that makes it difficult to recognise the characteristic stellar sequences. This scatter is mainly due to the different distance of the field stars and, to a lesser extent, to their different reddening.
Moreover, since the coordinates of these MSPs points at different directions in the Galaxy they sample different stellar populations, i.e.  from the Galactic Disk, Bulge, and Halo. In the case of the PSR\, J2017+0603 field, however, the observed CM diagram shows evidence of a more defined stellar sequence. This suggests that most stars in the field of view belong to a more homogeneous stellar population, e.g. in an open cluster, although there is no known open cluster (or candidate) in the pulsar field. Alternatively, they might all be at a comparable distance from the Sun,
as suggested by the line of sight to the pulsar field, which intercepts 
the Sagittarius spiral arm. 
For all pulsars, the location of the companion stars in the observed CC diagrams lies in most cases along, or close to the field stars, suggesting that the companions have no unusual colours. This might indicate that their colours would  not be very much affected by an hypothetical  irradiation from the MSP. 

%

 \begin{table}
\begin{center}
\caption{Pulsar dispersion measure   (DM), obtained from the ATNF pulsar data base, and distance (D), inferred from the Galactic free electron density along the line of sight  (Cordes \& Lazio 2002), of the last three MSPs listed  Table \ref{psr}.  For the distance, we assumed an uncertainty of $\pm$ 20\%. The fourth and fifth  columns give the hydrogen column density $N_{\rm H}$  derived from the fits to the MSP X-ray spectra (Abdo et al.\ 2013) and the Galactic extinction in the pulsar direction ($A_V$) derived from the $N_{\rm H}$   using the relation of Predehl\&Schmitt (1995).
}
\label{dist}
\begin{tabular}{lcccc} \hline
Pulsar 				 &  DM                    & D        & $N_{\rm H}$              & $A_V$ \\ 
                                           &   (pc cm$^{-3}$) & (kpc)  & ($10^{20}$ cm$^{-2}$) &  \\ \hline
J0614$-$3329$^1$	                  & 37.049                &	 1.88  & $6.44^{+6.32}_{-2.01}$ & $0.36^{+0.35}_{-0.12}$  \\
J1231$-$1411	 & 8.09 &	 0.43 &  $11.3\pm 5.1$ &  0.63$\pm$0.28\\
J2017+0603    	 & 23.918 &	 1.57 &  10&  0.56 \\ 
\hline
\end{tabular}
$^1$ The distance based on the DM might be overestimated by a factor of two or more (Ransom et al.\ 2011).
\end{center}
\end{table}


\section{Discussion}

We compared the positions of the candidate companion stars in the observed CM and CC diagrams with stellar population models and template WD evolutionary tracks
%
accounting for both the 
distance to the pulsar 
and the interstellar extinction along the line of sight.  
For none of the three MSPs there is a distance measurement based on the radio parallax. Thus, with all due caveats, we used as a reference the pulsar distance obtained from the dispersion measure   (DM) and the Galactic free electron density $n_e$ along the line of sight  (NE2001; Cordes \& Lazio 2002).  We assumed a realistic uncertainty of $\pm$20\% on the computed distance, as the authors recommended. This uncertainty is much larger than that derived from the errors on the DM but accounts, in most cases, for systematic effects related to the possible under/over estimation of the Galactic electron density in certain directions. For PSR\,  J0614$-$ 3329, however, the uncertainty on the distance is probably much larger than 20\% since the line of sight to  the pulsar is almost tangent to the Gum Nebula and the distance is likely half as estimated from the NE2001 model (Ransom et al.\ 2011).  
%
No direct measurement of the interstellar reddening along the line of sight is available for these pulsars. However, all of them have been detected in the X rays (Abdo et al.\ 2013). Therefore, as a reference, we estimated the reddening from the hydrogen column density $N_{\rm H}$ along the line of sight derived from the spectral fits to the X-ray spectra after applying the relation of Predehl \& Schmitt (1995). Then, we computed the extinction in the different filters using the extinction coefficients of Fitzpatrick (1999).  We note that for both PSR\, J0614$-$3329 and J2017+0603 the X-ray spectrum is poorly constrained (see Table 16 in  Abdo et al.\ 2013), and so is the value of the column density $N_{\rm H}$. In particular, for the PSR\, J2017+0603 the  $N_{\rm H}$ was set to the Galactic value in the pulsar direction and scaled for the pulsar distance ($10^{21}$ cm$^{-2}$). We obtained a new estimate of the $N_{\rm H}$ from the pulsar DM by applying the linear fit between these two quantities computed by He et al.\ (2013).  For the DM towards PSR\, J2017+0603 (23.918 pc cm$^{-3}$) the fit yields $N_{\rm H}= 7.17^{+3.11}_{-2.15} \times 10^{20}$ cm$^{-2}$, where the errors are associated with the fit 90\% confidence interval.  The $N_{\rm H}$ value computed from  the He et al.\ (2013) fit is consistent with that reported in Abdo et al.\ (2013) and corresponds to  $A_V= 0.40^{+0.17}_{-0.12}$, after applying the relation of Predehl \& Schmitt (1995).
The DM, inferred distance, $N_{\rm H}$, and estimated interstellar reddening along the line of sight for the three MSPs are summarised in Table \ref{dist}.

Firstly, we compared the observed CM and CC diagrams of the MSP fields with simulated stellar sequences computed from the Besan\c{c}on models (Robin et al.\ 2004). We simulated these sequences for different stellar populations, i.e belonging to the MS, Red Giant Branch (RGB), or the WD branch, and for distance values up  to 15 kpc. The simulated sequences for each MSP field are shown in  the panels in Figure \ref{0614-cmd}--\ref{2017-cmd} as the grey scale maps. 
In the CM diagrams,  the dark grey regions correspond to distance values within $\pm 20\%$  the assumed pulsar distance  (see Table \ref{dist}), whereas in the CC diagrams they correspond to magnitudes within $\pm$ 0.05 the r-band magnitude of the pulsar candidate companion star. 
 As seen, the spread in the simulated stellar sequences well reproduces the spread in the observed points, as expected for different stellar populations at different distances. Since the field stars are affected by an unknown interstellar extinction, and the simulations based on the Besan\c{c}on models simply compute a reddening scaled proportionally to the assumed distance in a given direction, introducing a reddening correction in our simulations might bias a direct comparison between the observed and the simulated stellar sequences.  Therefore, for simplicity, in all cases we simulated the stellar sequences assuming a null reddening. 
Each panel also shows the reddening vector corresponding to the E(B-V) estimated from  the $N_{\rm H}$ measured along the line of sight to each MSP (Table \ref{dist}) and computed using the extinction coefficients of Fitzpatrick (1999). Then, we used the reddening vectors as a reference to trace the extinction-corrected locations of the observed points for the MSP companion stars (blue points) along the simulated stellar sequences. 

As seen, for all the three MSPs the location of the candidate counterpart in the diagrams falls off the region of the simulated MS and close to that of the simulated WD sequence. This is consistent with a WD identification for all the candidate companion stars, as proposed in the literature (Ransom et al.\ 2011; Cognard et al.\ 2011).
 We note that non-degenerate  companion stars as those in BW or RB systems would be much closer than observed to the late MS (see, e.g. Pallanca et al.\ 2012). Moreover, 
 the minimum mass of the companion stars and the orbital period of the binary system (Table \ref{orb})
would rule out that these MSPs are BWs or RBs (see, e.g. Figure 1 in Roberts 2013).  Indeed, the minimum companion masses are larger than those of BW companions, whereas the orbital periods are larger than those of known RB binary systems, which have all orbital periods  shorter than 1 d. The latter case is especially true for $\0614$ which has an orbital period of  53.584 d. 

\begin{figure*}
\begin{center}
\includegraphics[width=160mm]{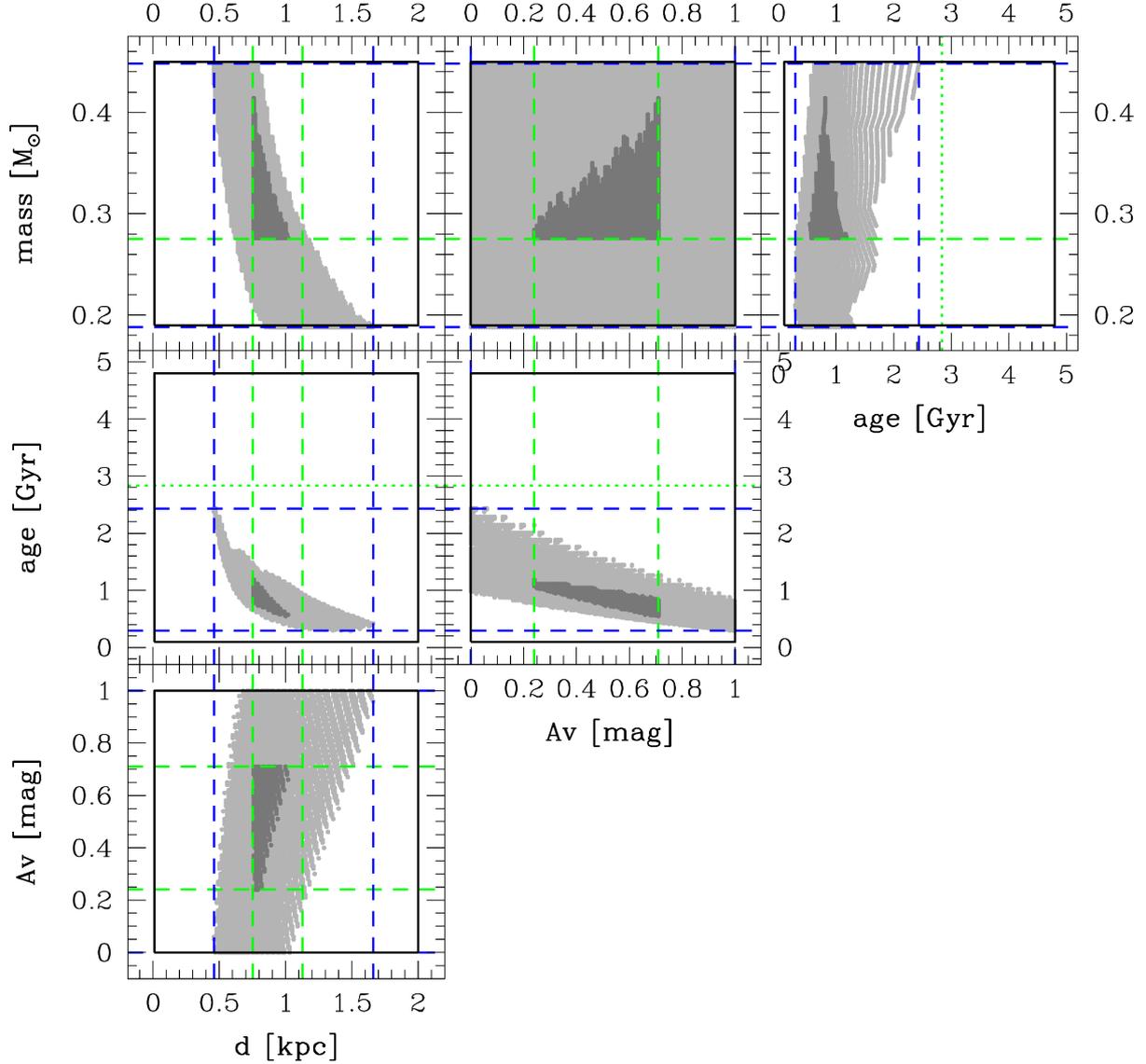}
  \caption{Pair combinations of the four parameters which we used to fit the observed magnitudes and colours of the candidate companion  star to $\0614$ with mode evolutionary tracks (Panei et al.\ 2007). This plot refers to the case of an He WD}. The thick black lines mark the investigated region in each plane of the parameter space. The light grey regions mark, for each pair of parameters, the allowed parameter configurations which 
reproduce the observed optical properties of the candidate companion stars (magnitudes, colours). The extreme of these regions projected on each axis are marked by the blue lines, while the green dashed lines mark the parameter range constrained by the radio and X-ray measurements. Although the radio/X-ray constraints of each parameter pair seem to be more stringent than the optical ones, imposing that all conditions are simultaneously satisfied  shrinks the allowed parameter configurations to the dark the grey regions. The dotted green line corresponds to the pulsar spin-down age. 
\label{limiti0614}
\end{center}
\end{figure*}

\begin{figure*}
\begin{center}
\includegraphics[width=160mm]{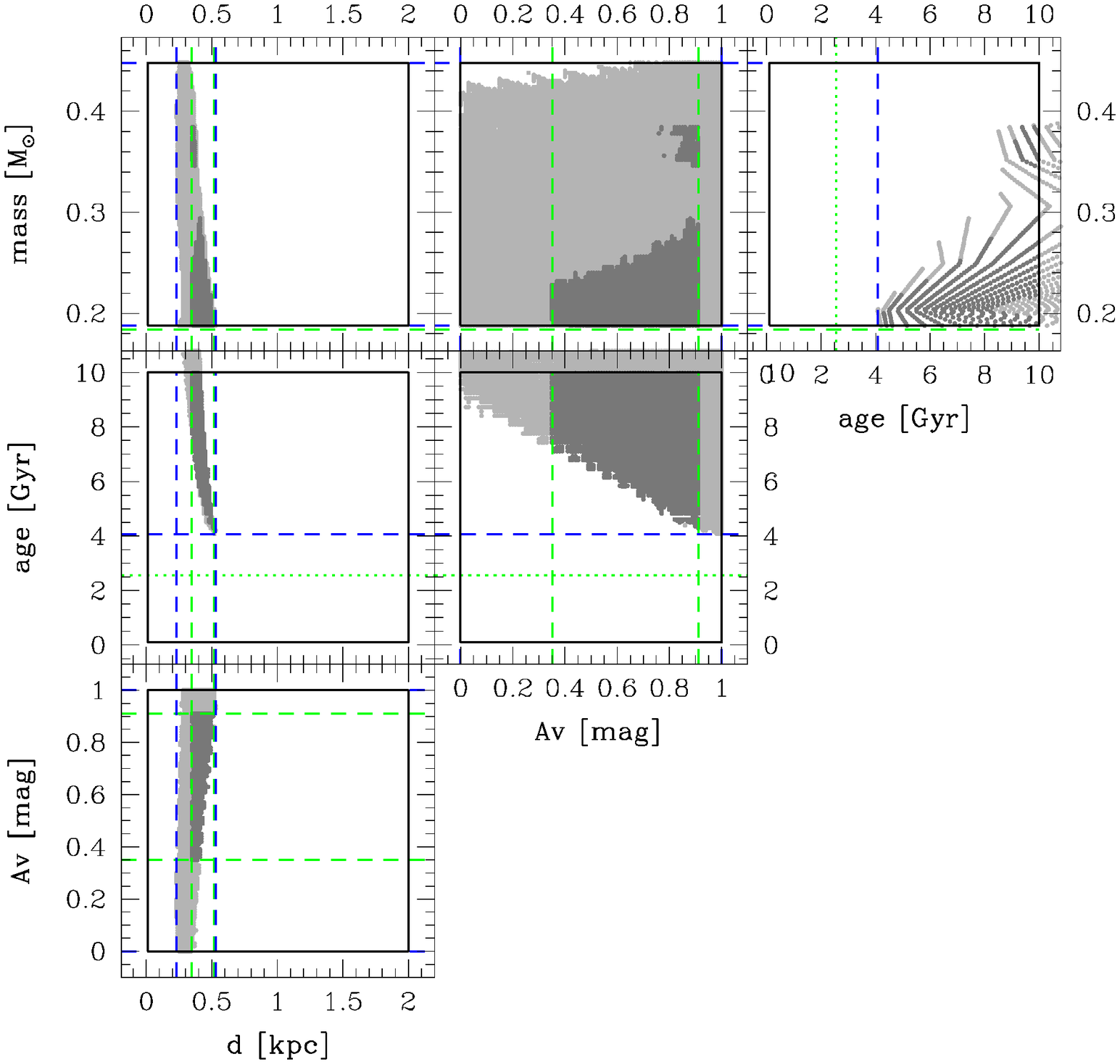}
  \caption{Same as Fig. \ref{limiti0614} but for $\1231$.   }\label{limiti1231}
\end{center}
\end{figure*}

\begin{figure*}
\begin{center}
\includegraphics[width=160mm]{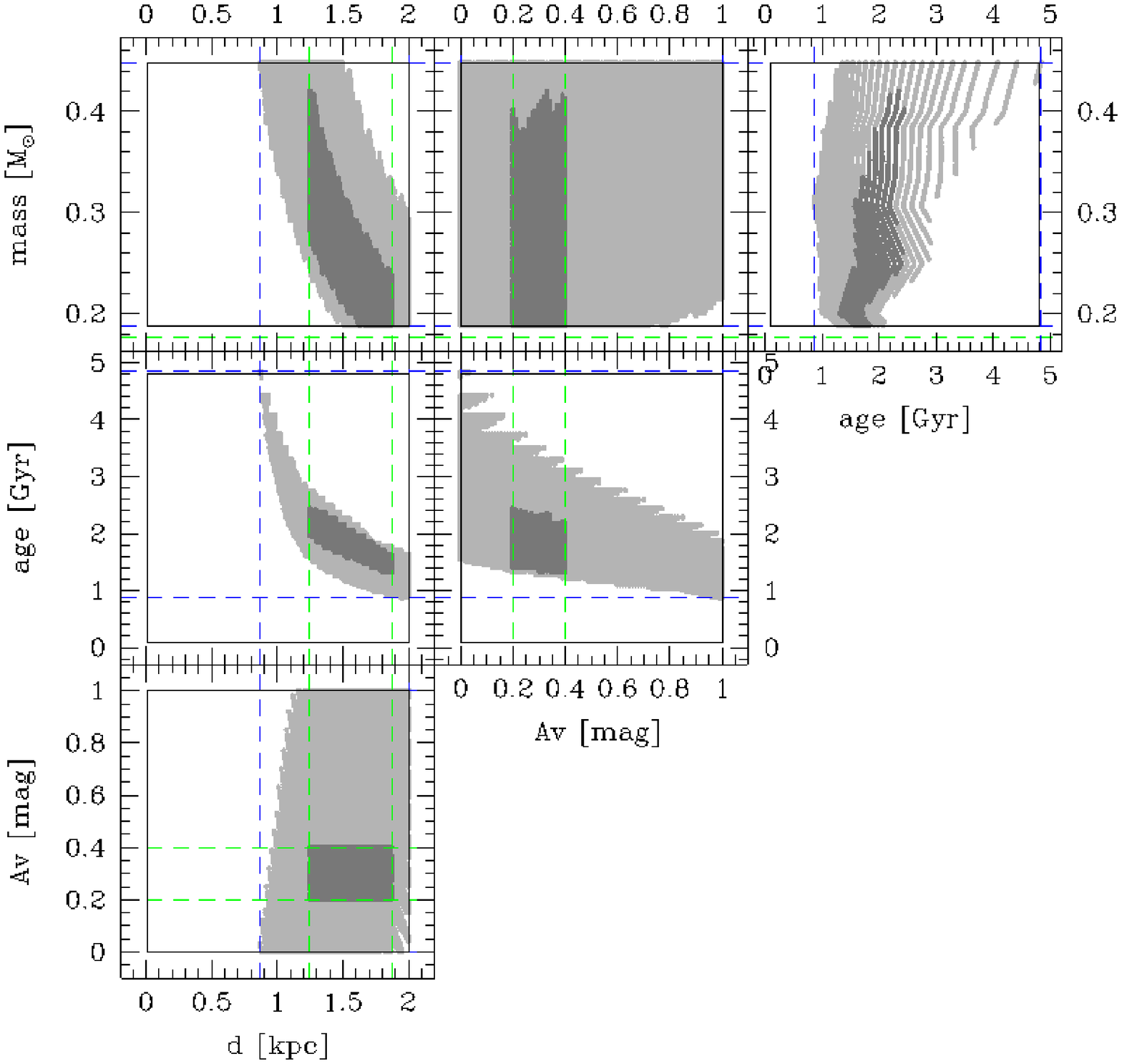}
  \caption{Same as Fig. \ref{limiti0614} but for  $\2017$.  }\label{limiti2017}
\end{center}
\end{figure*}

In order to better determine  the WD characteristics, obtain a first tentative estimate of their mass and an independent estimate of the age of the binary system, we used the model evolutionary tracks of Panei et al.\ (2007), which are computed for both He and low-mass CO WD types, and different WD masses. In particular, these tracks are quite suited to our goal since the assumed range of WD masses is close to the minimum mass of the companions inferred from the mass function of the binary systems (Table \ref{orb}). These tracks\footnote{{\tt http://fcaglp.fcaglp.unlp.edu.ar/evolgroup/TRACKS/\\tracks\_heliumcore.html}} also predict magnitude values for both WD types in the same photometric system as used by the G-MOS observations, i.e. the SDSS one (Fukugita et al.\ 1996).  
 As an example,  Figure \ref{0614-cmd}--\ref{2017-cmd} show the evolutionary tracks simulated for both an He WD  (light blue) and a low-mass CO WD (dark blue).
  
 The evolutionary tracks span an age range up to 4.8 Gyr, where the age limit corresponds to that of the available model (Panei et al.\ 2007), and different masses.
 As a reference, we also plotted the evolutionary tracks of MS  stars  for both Z=0.02 (red solid lines) and  Z=0.0001 (red dashed lines)  and for different mass ranges  (0.5--1 $M_{\odot}$, with steps of  0.1 $M_{\odot}$). 
 In each figure, the lines correspond to the
 nominal value of the MSP distance (Table \ref{dist}).
 Following Ransom et al.\ (2011), for PSR\, J0614$-$3329 we assumed a distance value half of that obtained from the DM and the NE2001 model.  As done for the simulated stellar sequences computed from the Besan\c{c}on models, we plotted the simulated WD and MS evolutionary tracks assuming a null extinction, for simplicity.  As seen, for both  PSR\, J0614$-$3329 and J2017+0603 the location of the candidate companion star in the CM and CC diagrams  seems to be more compatible with the simulated evolutionary tracks for an He WD (light blue lines), 
than for a low-mass CO WD (dark blue lines). 
 The difference between the He and CO WD evolutionary tracks is more evident for PSR\, J0614$-$3329, whereas
 for J2017+0603  the r-i colour is partially compatible with both WD types.
   The location of the candidate companion star to PSR\, J1231$-$1411 in the CM and CC diagrams does not overlap with any of the evolutionary tracks, although it seems to be consistent with the extrapolation of the He WD evolutionary tracks. Since these are computed for a maximum age of 4.8 Gyr, it means that the companion star must be significantly older than those of PSR\, J0614$-$3329 and J2017+0603.  
  To summarise, our qualitative analysis of the CM and CC diagrams 
  seems to suggest that the candidate companions to these three MSPs are  more probably He WD than low-mass CO WDs.
However, owing to the uncertainties in extrapolating the evolutionary tracks for ages above 4.8 Gyr, we regard the possible identification of the  PSR\, J1231$-$1411  companion as an He WD as more uncertain than in the other cases. In addition, it should be considered that CO WDs tracks tend to separate and spread out at low effective temperatures due to different atmopheric compositions. An example of this can be seen in Bergeron et al.\ (2011) and refs. therein.

We  tried to verify more quantitatively a possible association with either of the two types for different 
values of
the WD mass and age, and for different values of the 
distance and reddening. To this aim, we systematically investigated all possible combinations in a four dimensional parameter grid  and selected only those for which the optical magnitudes and colours predicted by the model evolutionary tracks for He or low-mass CO WDs were found to be consistent with the observed ones,  within the measured photometry uncertainties. We considered the case of He WDs first.
 Fig. \ref{limiti0614},   \ref{limiti1231} and  \ref{limiti2017} show  the allowed combinations for each of the selected pairs of parameters for the He WD case. 
For visualisation purposes, in each panel  the regions corresponding to the allowed parameter combinations are plotted in light grey.
The discontinuity in the grey regions noticed in some panels are purely an effect of the parameter quantisation. The black thick lines in each panel mark the investigated parameter range. We considered a range of values around those expected from radio and optical observations, with a generous tolerance to account for the associated uncertainties. In particular, we considered a range of distances $0.01<d<2$ kpc (0.01 kpc steps) and $0<A_{\rm V}<1$ (0.01 magnitude steps). For the companion mass we considered the range $0.188 M_{\odot} <\mcom <0.488 M_{\odot}$ (0.002 $\Msun$ steps), which is that of the model evolutionary track for He WDs (Panei et al.\ 2007). 
We derived WD tracks for intermediate mass values from a linear interpolation between the tracks corresponding to the available mass values.  Finally, for the age we considered a  range of values obtained from the computed WD tracks, for a given value of WD  mass and extinction, with steps of 0.1 magnitudes in the r-band magnitude.   We remind that for PSR\, J1231$-$1411  the comparison with the WD tracks is based on  their extrapolation for ages above 4.8 Gyr and is obviously more uncertain than for the other two MSPs. 

%
Possible parameter ranges, purely imposed by the observed colours and magnitudes of the companion star, are represented by the blue dashed lines intercepting tangentially the light grey regions in each panel in Fig. \ref{limiti0614}--\ref{limiti2017}.  Note that in some panels the black thick lines and the dashed blue lines coincide. This corresponds to those cases where the uncertainty on our photometry is not sufficiently small to better constrain the parameter value.  Then,
we considered, with the due caveats, the limits imposed by the radio observations on the MSP parameters, such as the lower limit on the companion mass $M_{\rm C}$ (Table \ref{orb}) and the pulsar distance (Table \ref{dist}), and the limits imposed by the X-ray observations, such as the $N_{\rm H}$ and the inferred interstellar extinction (Table \ref{dist}). 
We did not apply any limit to the WD age based on the pulsar spin-down age because of the well known difficulties in determining a reliable uncertainty range for this value other than that the formal error derived from the measured $P_{\rm s}$ and $\dot{P_{\rm s}}$, of course (e.g., Lorimer \& Kramer 2005). 
Finally, we selected only the configurations for which all conditions imposed to all pair of parameters by the limits derived from optical, radio, and X-ray observations were simultaneously satisfied.  The selected parameter configurations are represented by the dark grey regions in the panels in Fig. \ref{limiti0614}--\ref{limiti2017}.  For $\1231$, the obtained age range originally spanned values larger than the age of the Universe. Therefore, we  imposed that the WD age is smaller than an arbitrary value, which we set to 10 Gyr.  The plots in Fig. \ref{limiti1231} were updated accordingly.

We repeated 
the same analysis as above for the  low-mass CO WD case. We assumed the same range of distance and  extinction values as in the previous case and the range of companion star masses ($0.351 M_{\odot} < M_{\rm C} < 0.448 M_{\odot}$)  from the model evolutionary tracks for low-mass CO WDs (Panei et al.\ 2007).   We computed the age range exactly as in the previous case.  We found that all the allowed configurations (not shown here) imply companion masses of $\la$0.4 $M_{\odot}$ ($\0614$, $\2017$) and $\la$0.36 $M_{\odot}$ ($\1231$), i.e. at the low end of the mass range even for low-mass CO WDs. This would make a CO WD companion for the three MSPs somewhat less likely, although this possibility can be firmly ruled out only with a better characterisation of the star spectra through follow-up spectroscopy observations. 
 Our results are in line with the qualitative analysis of the CM  and CC diagrams (Figure \ref{0614-cmd}--\ref{2017-cmd}) and the possibility that the candidate companion stars of the three MSPs are He WDs. 


The derived parameter ranges for the three MSP companions 
are summarised in Table \ref{tab:parameters}, for the He WD case. In the first half, we report the ranges of WD age, mass, distance, and extinction obtained from the above analysis (Fig. \ref{limiti0614}--\ref{limiti2017}), whereas in the second half we report the values of the WD surface temperature, luminosity and surface gravity extracted from the model, which correspond to the obtained age and mass ranges.
Unfortunately, for both $\1231$ and PSR\, J2017+0603, which are the MSPs with the faintest companion stars (Table \ref{stars}), the photometry errors were so large that several sets of parameters were able to fit the observed optical magnitudes and colours, although the parameter degeneracy has decreased (see Fig. \ref{limiti1231} and  \ref{limiti2017}).  On the other hand, the smaller photometry errors for $\0614$ made it easier to find possible ranges for the different parameters.   We note that our results are influenced by the   model uncertainties, including the characterisation of the WD atmosphere and composition, and by the overall uncertainties on the predicted magnitudes, which are more difficult to quantify than the  photometry errors on the observed magnitudes, by some arbitrary assumptions on the parameter uncertainties, such as on the pulsar distance, 
and by the uncertainty on the extrapolated parameters, such as the interstellar extinction.
Therefore, the inferred ranges for the WD parameters should not be taken rigidly but only as indicative.   This is particularly true in the case of $\1231$ (see above).
Spectroscopy observations and multi-band light curves will be needed to obtain a better characterisation of the companion star properties and pave the way to  a more robust comparison with stellar models. 

\begin{figure*}
\centering
{\includegraphics[height=8.5cm,angle=270,clip=]{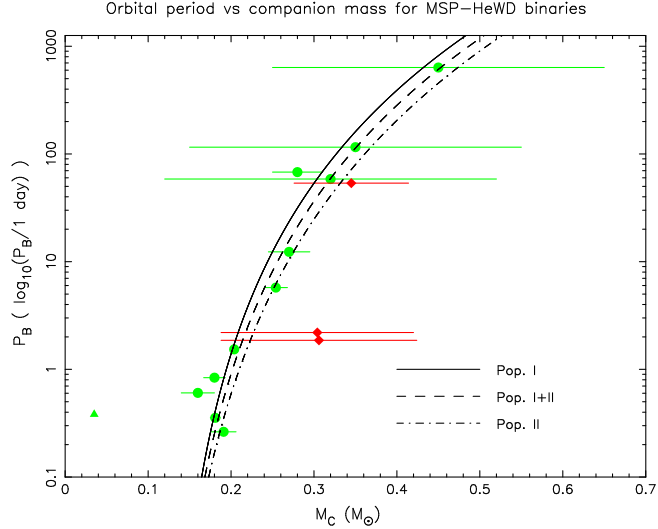}}	
\caption{\label{pb_mc}  Orbital period $P_{\rm b}$ vs. companion mass $M_{\rm C}$ for MSP-He-WD binaries (green filled circles; updated from Corongiu et al.\ 2012).  Error bars correspond to $1\sigma$ uncertainties in the masses. If not visible, their size is smaller than the symbol. The red diamonds correspond to the three MSPs for which we identified the companion stars:  $\0614$, $\2017$, and $\1231$ (top to bottom). The black filled triangle corresponds to the BW PSR\, B1957+20. The solid, dashed, and dot-dashed lines are the tracks corresponding to the theoretical correlation found by Tauris \& Savonije (1999).}
\end{figure*}

For all our pulsars, we found a possible range of values for  the companion mass $M_{\rm C}$.
As it has already been observed,  a correlation has been found between the orbital period $P_{\rm b}$ and the companion mass $M_{\rm C}$ for binary systems hosting a MSP and an He WD (see, e.g. Corongiu et al.\  2012 and references therein). Such a correlation was theoretically found by Tauris \& Savonije (1999) and has been confirmed by the mass measurements of the companion star in about a dozen of such systems. The predictions by Tauris \& Savonije (1999) indicate, for the three MSP binary systems discussed here, a value for the companion mass that would lie in the range that we obtained from our estimates (Table \ref{tab:parameters}). This is seen in Fig. \ref{pb_mc} (updated from Corongiu et al.\ 2012), where we show the $P_{\rm b}$-$M_{\rm C}$ plot for all MSPs with an He WD companion of measured mass  together with the estimated masses for the companion stars to  $\0614$, $\1231$, and $\2017$.

For both $\0614$ and  PSR\, J2017+0603, the inferred age estimate for the WD companion  would be smaller than the pulsar spin-down age $\tau$  by a factor of $\sim$2.4--5 (Table \ref{psr}). On the other hand, in the case of $\1231$ the minimum age of the WD companion obtained from our analysis would be  $\sim$ 70\% greater than the  pulsar spin-down age.   However, it has been shown (e.g., Tauris 2012) that spin-down ages are not reliable age estimators for MSPs. Therefore, we must take this discrepancy with the due care.
Since for $\1231$ we had to impose an arbitrary upper limit to the age of the companion star, only for $\0614$ and $\2017$ we could obtain a first estimate of the cooling age of the WD, which  would be of the order of about 0.8 and 1.7 Gyr, respectively.

\begin{table*}
 \begin{center}
  \begin{tabular}{lccc}
  \hline
                              &      PSR\, J0614$-$3329             &      PSR\, J1231$-$1411                  &     PSR\, J2017+0603\\
 \hline    
   $\mcom$ $[\Msun]$    &     0.276--0.414    &       0.188--0.384      &     0.188--0.42\\
     Age [Gyr]          &     0.55--1.18                   &       4.30--10                       &    1.31--2.44\\
    D [kpc]              &      0.76--1.02                 &       0.35--0.51                       &     1.25--1.87\\
    $\Av$  [mag]    &     0.24--0.71              &       0.35--0.91                 &     0.2--0.4\\
    \hline
    $T$ [K]		         & 7237--9666		&	2652--4009		&  5412--6437\\
    $log ~[L/L_{\odot}]$  &   [-2.898, -2.386]    &	[-4.549, -3.920]			&    [-3.346, -3.014]         \\
    $log ~g$		&7.107--7.572		&	6.973--7.603			&  6.778--7.625\\ 
   \hline
    \end{tabular}  
    \caption{Parameters for the candidate companion stars to the MSPs obtained from Fig. \ref{limiti0614}--\ref{limiti2017} and the comparison with He WD model evolutionary tracks (Panei et al.\ 2007). Top section:  derived ranges for the mass and age of the companion star ($\mcom$) to the MSP,  distance (D) and interstellar extinction along the line of sight ($\Av$).    These ranges basically correspond to the outermost values of the dark grey regions of Figures \ref{limiti0614}, \ref{limiti1231} and \ref{limiti2017}.
    In  case of $\1231$, 
    these ranges were obtained by imposing that the WD age is $<10$ Gyr. 
    Bottom section: ranges of WD surface temperature $T$, luminosity $L$, and gravity $ log~ g$ corresponding to the above mass and age ranges.}
       \label{tab:parameters}
      \end{center}
\end{table*}

For 
$\0614$ the derived range for the pulsar distance
purely imposed by the observed colour and magnitude of the companion star (blue lines in Fig. \ref{limiti0614}), would confirm that the distance to the pulsar is indeed about half the value inferred from the DM and the NE2011 model (Table \ref{dist}), 
as suggested by Ransom et al.\ (2011).  
On the contrary, the corresponding range for $\1231$ would indicate that the pulsar distance cannot be much larger than the value obtained from the DM, whereas  its high Galactic latitude ($b \sim 48\degr$) and the low value of the DM (8.09 pc cm$^{-3}$) might have suggested that such a value is underestimated by a factor of two (Gaensler et al.\ 2008; Chatterjee et al.\  2009).


We  used the mass and age ranges of the WD companions  derived above to select the associated ranges of  the star surface temperature $T$, luminosity and surface gravity $log~g$ (second half of Table \ref{tab:parameters}) from the He WD model evolutionary tracks (Panei et al.\ 2007).  As seen, the WD companion star to $\1231$  would stand out for its lower temperature ($T\sim 2600$--4000 K), definitely on the lower end of the WD temperature distribution,  and lower luminosity  ($log ~[L/L_{\odot}]\approx -4$) with respect to the other MSP companions. The inferred lower temperature was also anticipated by the observed reddish colour of the star (Fig. \ref{1231-cmd}). Thus, the WD companion to $\1231$  would be one of the coolest WDs known so far.   If confirmed, this result  would be exceptional but not unheard of. We note that a very low temperature WD companion ($T<3000$ K) has been likely identified also in the MSP binary system PSR\, J2222$-$0137  (Kaplan et al.\ 2014). On the other hand,  both the WD companion stars to $\0614$ and $\2017$ would be hotter, with temperatures more compatible with the values expected for a WD ($T\sim 7000$--10000 K), and are more luminous  ($log ~[L/L_{\odot}]\approx -3$).

\section{Summary and conclusions}

Using the Gemini-South telescope, we identified the likely companion stars to three MSPs (PSR\, J0614$-$3329, J1231$-$1411, and J2017+0603).
For a fourth pulsar (PSR\, J0613$-$0200), 
the identification of the companion star was hampered by the presence of a bright star (g=16$\pm$0.03) at $\sim 2\arcsec$ from the pulsar radio position.  We ruled out with a reasonable confidence that this star is a candidate companion to the pulsar. The companion stars of the other three MSPs  can be possibly identified as He WDs on the basis of multi-band photometry and comparison with simulated stellar models (Panei et al.\ 2007).  For PSR\, J1231$-$1411 the identification is somewhat more uncertain owing to the more ambiguous match with the evolutionary tracks.  If these identifications were confirmed, they would prove that He WDs tend to be the most common companions to non-eclipsing MSPs, as suggested by the number of MSP companions which have been identified in the optical so far.  For PSR\, J0614$-$3329, J1231$-$1411, and J2017+0603, we  derive possible ranges for the WD mass, age, surface temperature, luminosity, and gravity (see Table \ref{tab:parameters}) within the ranges predicted by the assumed He WD models and optical/radio observations.  
For none of these three MSPs we could look for flux modulations at the orbital period of the binary system, owing to the sparse data points and the uneven time coverage of the orbital period provided by the Gemini observations.  Future variability studies of the companion star fluxes through optical photometry, together with the measurements of the radial velocity curves through optical spectroscopy with 8m-class telescopes, will firmly prove the proposed associations with the MSPs, currently based upon a very good positional coincidence with the radio coordinates and the colours of the candidate companions stars. 
Optical spectroscopy would also be crucial to   verify our tentative classification of the MSP companion stars as He WDs, so far only based upon optical broad-band photometry in three bands, obtain more robust measurements of their stellar parameters, such as the mass, surface temperature and gravity,  and provide information on the atmosphere composition.

\section*{Acknowledgments}
We thank the anonymous referee for his/her constructive comments to the manuscript.
We are grateful to Dr. Achille Nucita (Dept. of Mathematics and Physics, University of Salento) for his advice in the variability analysis. The research leading to these results has received funding from the European Commission 
Seventh Framework Programme (FP7/2007-2013) under grant agreement n. 267251. This research is part of the project COSMIC-LAB funded by the European Research Council (under contract ERC-2010-AdG-267675). This research is based on observations obtained at the Gemini Observatory, which is operated by the 
Association of Universities for Research in Astronomy, Inc., under a cooperative agreement 
with the NSF on behalf of the Gemini partnership: the National Science Foundation 
(United States), the National Research Council (Canada), CONICYT (Chile), the Australian 
Research Council (Australia), Minist\'{e}rio da Ci\^{e}ncia, Tecnologia e Inova\c{c}\~{a}o 
(Brazil) and Ministerio de Ciencia, Tecnolog\'{i}a e Innovaci\'{o}n Productiva (Argentina).
This research has made use of the APASS database, located at the AAVSO web site. Funding for APASS has been provided by the Robert Martin Ayers Sciences Fund.

\label{lastpage}


\begin{thebibliography}{99}


\bibitem[\protect\citeauthoryear{Abdo et al.}{2009}]{abdo09} Abdo A. A., et al.,  2009, Science, 325, 848  


\bibitem[\protect\citeauthoryear{Abdo et al.}{2013}]{abdo13} Abdo A. A., et al.,  2013, ApJS, 208, 17 

\bibitem[Alpar et al.(1982)]{alpar82} Alpar M.~A., Cheng A.~F., Ruderman M.~A., \& Shaham J.\ 1982, Nature, 300, 728 


\bibitem[Benvenuto et al.(2014)]{ben14} Benvenuto O. G., De Vito M. A., Horvath J. E., 2014, ApJ, 786, L7

\bibitem[Bergeron et al.(2011)]{ber11} Bergeron P., et al., 2011, ApJ, 737,28

\bibitem[\protect\citeauthoryear{Bertin \& Arnouts}{1996}]{bs96}Bertin E. \& Arnouts S., 1996, A\& A Suppl., 117, 393 

\bibitem[Bhattacharya \& van den Heuvel(1991)]{bhattvan91} Bhattacharya, D., \& van den Heuvel, E.~P.~J.\ 1991, \physrep, 203, 1 

\bibitem[\protect\citeauthoryear{Chatterjee et al.}{2009}]{cog11}Chatterjee S., et al., 2009, ApJ, 698, 250

\bibitem[\protect\citeauthoryear{Cognard et al.}{2011}]{cog11}Cognard I., et al., 2011, ApJ, 732, 47 

\bibitem[\protect\citeauthoryear{Collins et al.}{2011}]{col11}Collins S., Shearer A., Mignani R. P., 2011, in Radio pulsars: an astrophysical key to
unlock the secretes of the universe, AIP Conf. Proc., 1357, 310 

\bibitem[\protect\citeauthoryear{Cordes \& Lazio}{2002}]{2002astro.ph..7156C} Cordes J.~M., \& Lazio T.~J.~W., 2002, arXiv:astro-ph/0207156 

\bibitem[\protect\citeauthoryear{Corongiu et al.}{2012}]{cor12} 	Corongiu A., et al., 2012, ApJ, 760, 100

\bibitem[\protect\citeauthoryear{Fitzpatrick}{1999}]{1999PASP..111...63F} Fitzpatrick E. L., 1999, PASP, 111, 63
	
	
\bibitem[\protect\citeauthoryear{Fukugita et al.}{1996}]{fuk96} Fukugita M., Ichikawa T.,  Gunn J. E.,  Doi M., Shimasaku K.,  Schneider D. P., 1996, AJ, 111, 1748 

\bibitem[\protect\citeauthoryear{Gaensler et al.}{2008}]{gae08}Gaensler B. M., Madsen G. J., Chatterjee S., \& Mao S. A., 2008, PASA, 25,184

 
\bibitem[\protect\citeauthoryear{Henden \& Munari}{2014}]{hen14} Henden A. \& Munari U., 2014, Contributions of the Astronomical Observatory Skalnat\'e Pleso, vol. 43, no. 3, p. 518 

\bibitem[\protect\citeauthoryear{He et al.}{2013}]{he13} He, C., Ng, C.-Y., Kaspi ,V. M., 2013, ApJ, 768, 64

\bibitem[\protect\citeauthoryear{Kaplan et al.}{2014}]{kap14} Kaplan D. L., et al., 2014, ApJ, 789, 119


\bibitem[\protect\citeauthoryear{Lasker et al.}{2008}]{las08}  Lasker B. M., et al., 2008, AJ,136, 735 

\bibitem[\protect\citeauthoryear{Lattanzi et al.}{1997}]{1997A&A...318..997L} Lattanzi M.~G., Capetti A., \& Macchetto F.~D., 1997, A\&A, 318, 997 


\bibitem[\protect\citeauthoryear{Lauberts \& Valentijn}{1989}]{} Lauberts A. \& Valentijn E. A., 1989, The Surface Photometry Catalogue of the ESO-Uppsala Galaxies (Garching: European Southern Observatory)


 \bibitem[\protect\citeauthoryear{Lorimer et al.}{1995}]{lor95}  Lorimer D. R., et al.,  1995, ApJ, 439, 933 
 
\bibitem[Lorimer \& Kramer (2005)]{handbook} Lorimer, D.~R., Kramer, M.\ 2005, {\it Handbook of pulsar astronomy}, Cambridge University Press  
 
\bibitem[\protect\citeauthoryear{Lyne et al.}{1987}]{lyne87} Lyne, A.~G., Brinklow, A.,   Middleditch, J., Kulkarni, S.~R., \& Backer, D.~C.\ 1987, \nat, 328,   399 

\bibitem[\protect\citeauthoryear{Manchester et al.}{1996}]{man96} Manchester R. N., et al., 1996, MNRAS, 279, 1235 

\bibitem[\protect\citeauthoryear{Manchester et al.}{2005}]{man05} Manchester R. N., Hobbs G. B., Teoh A. \& Hobbs M., 2005, AJ, 129, 1993 

\bibitem[\protect\citeauthoryear{Mignani et al.}{2014}]{}Mignani R.P., et al.,  2014, MNRAS,  443, 2223 

\bibitem[\protect\citeauthoryear{Mucciarelli et al.}{2013}]{} Mucciarelli A., Salaris M., Lanzoni B., Pallanca C., Dalessandro E., Ferraro F. R., 2013, ApJ, 772, L27

\bibitem[\protect\citeauthoryear{Oke}{1974}]{oke74}  Oke J.B., 1974, ApJS, 27, 21 
 
\bibitem[\protect\citeauthoryear{Orosz \& van Kerkwijk}{2003}]{oro03}  Orosz J. A., van Kerkwijk M. H.,  2003, A\&A, 397, 237
 
\bibitem[\protect\citeauthoryear{Pallanca et al.}{2012}]{}Pallanca C., Mignani R. P.,  Dalessandro E., Ferraro F. R.,  Lanzoni B., Possenti A.,  Burgay M.,  Sabbi E., 2012, ApJ, 755, 180

\bibitem[\protect\citeauthoryear{Pallanca et al.}{2013}]{}Pallanca C., Lanzoni B., Ferraro F. R., Dalessandro E., Possenti A., Salaris M., Burgay M., 2013, ApJ, 773, 127

\bibitem[\protect\citeauthoryear{Pallanca et al.}{2014}]{}Pallanca C., Ransom S. M., Ferraro F. R., Dalessandro E., Lanzoni B., Hessels J. W. T., Stairs I., Freire P. C. C., 2014, ApJ, 795, 29

\bibitem[\protect\citeauthoryear{Panei et al.}{2007}]{} Panei J. A., Althaus L. G., Chen X., Han Z., 2007, MNRAS, 382, 779


\bibitem[\protect\citeauthoryear{Possenti \& Burgay}{2008}]{pb08} Possenti A. \& Burgay M., 2008, in A decade of accreting millisecond X-ray pulsars, AIP Conf. Proc., Vol. 1068, p. 17  

\bibitem[\protect\citeauthoryear{Possenti}{2013}]{pos13} Possenti A., 2013, IAU Proc., Vol. 291, p. 121 

\bibitem[\protect\citeauthoryear{Predehl, P. \& Schmitt}{1995}]{ps95} Predehl P. \& Schmitt J.H.M.M. 1995, A\&A, 293, 889 

\bibitem[\protect\citeauthoryear{Ransom et al.}{2011}]{ran11} Ransom S. M., et al.,  2011, ApJ, 727, L16 


\bibitem[\protect\citeauthoryear{Roberts}{2013}]{rob13} Roberts M. S. E., 2013, Proceedings of the International Astronomical Union, Volume 291, pp. 127

\bibitem[\protect\citeauthoryear{Robin et al.}{2004}]{rob04} Robin A.C.,  Reyl\'e C., Derri\'ere S., Picaud S., 2004, A\&A 416, 157





\bibitem[\protect\citeauthoryear{Skrutskie et al.}{2006}]{skr06}  Skrutskie M.~F., et al.\ 2006, AJ, 131, 1163  

\bibitem[\protect\citeauthoryear{Smith et al.}{2007}]{smi07} Smith J. A., Allam S. S., Tucker D. L., Stute J. L., Rodgers C. T.,  Stoughton, C., 2007, AJ, submitted, under revision


\bibitem[\protect\citeauthoryear{Stappers et al.}{2001}]{sta01}  Stappers B. W., van Kerkwijk M. H., Bell J. F., Kulkarni, S. R.,  2001, ApJ, 548, L183

\bibitem[\protect\citeauthoryear{Stetson}{1987}]{ste87} Stetson P.B., 1987, PASP, 99, 191 

\bibitem[\protect\citeauthoryear{Stetson}{1994}]{ste94} Stetson P.B., 1994, PASP, 106, 250 

 \bibitem[\protect\citeauthoryear{Tauris \& Savonije}{1999}]{tau99}  Tauris T. M., \& Savonije G. J. 1999, A\&A, 350, 928
 
   \bibitem[\protect\citeauthoryear{Tauris}{2011}]{tau2011} Tauris T. M., 2011, in Evolution of compact binaries, ASP Conference Proceedings,  447, 285
   
  \bibitem[\protect\citeauthoryear{Tauris}{2012}]{tau2012}  Tauris T. M., 2012, Science, 335, 561  
  
   \bibitem[\protect\citeauthoryear{Tauris et al.}{2012}]{tau+2012} Tauris T. M., Langer N., Kramer M., 2012, MNRAS, 425, 1601
   
    \bibitem[\protect\citeauthoryear{Tauris}{2015}]{tau2015} Tauris T. M., Habilitation thesis, University of Bonn, 2015, arXiv:1501.03882
    
 \bibitem[\protect\citeauthoryear{van Kerkwijk}{2005}]{vdk05}      van Kerkwijk M.H., 2005, Binary Radio Pulsars, eds. F. A. Rasio \& I. H. Stairs, ASP Conf. Series, Vol. 328, p.357 

 
 \bibitem[\protect\citeauthoryear{Verbiest et al.}{2009}]{ver09} Verbiest J. P. W., et al., 2009, MNRAS, 400, 951
 
\bibitem[\protect\citeauthoryear{Zacharias et al.}{2013}]{zac13}   Zacharias N., Finch C. T., Girard T. M., Henden  A., Bartlett J. L., Monet D. G., Zacharias M. I.,  2013, AJ, 145, 44


\end{thebibliography}
\end{document}